\documentclass[journal]{IEEEtran}
\usepackage{amsmath,amsfonts}
\usepackage{algorithmic}
\usepackage{algorithm}
\usepackage{array}
\usepackage[caption=false,font=normalsize,labelfont=sf,textfont=sf]{subfig}
\usepackage{textcomp}
\usepackage{stfloats}
\usepackage{url}
\usepackage{verbatim}
\usepackage{graphicx}
\usepackage{cite}
\usepackage{multirow}
\usepackage{amsthm}

\theoremstyle{definition}
\newtheorem{definition}{Definition}[section]

\begin{document}

\title{Errorless Robust JPEG Steganography\\ using Outputs of JPEG Coders}

\author{
    \IEEEauthorblockN{Jan Butora, Pauline Puteaux and Patrick Bas}
}



\maketitle

\begin{abstract}
Robust steganography is a technique of hiding secret messages in images so that the message can be recovered after additional image processing. One of the most popular processing operations is JPEG recompression. Unfortunately, most of today's steganographic methods addressing this issue only provide a probabilistic guarantee of recovering the secret and are consequently not errorless. That is unacceptable since even a single unexpected change can make the whole message unreadable if it is encrypted. 
We propose to create a robust set of DCT coefficients by inspecting their behavior during recompression, which requires access to the targeted JPEG compressor. This is done by dividing the DCT coefficients into 64 non-overlapping lattices because one embedding change can potentially affect many other coefficients from the same DCT block during recompression. The robustness is then combined with standard steganographic costs creating a lattice embedding scheme robust against JPEG recompression. Through experiments, we show that the size of the robust set and the scheme's security depends on the ordering of lattices during embedding. We verify the validity of the proposed method with three typical JPEG compressors and the {\it Slack} instant messaging application. We benchmark its security for various embedding payloads, three different ways of ordering the lattices, and a range of Quality Factors. Finally, this method is errorless by construction, meaning the embedded message will always be readable.
\end{abstract}

\begin{IEEEkeywords}
robust steganography, recompression, lattice embedding, JPEG
\end{IEEEkeywords}
\section{Introduction}

With the wide usage of social networks and sharing platforms, the classical setup of steganography, which implies a lossless channel between Alice (the steganographer who embeds a payload) and Bob (the steganographer who decodes a payload), is meaningless in a lot of practical scenarios. 
This is due to the fact that the transmission channel involves a transcoding of the stego content (a JPEG recompression, for example) which can be seen as a noisy channel between Alice and Bob. 
In such a case, the errorless decoding of the payload is not possible anymore if Alice uses classical embedding schemes designed for lossless transmission (\textit{e.g.} in the JPEG domain with the use of J-Uniward~\cite{holub2014universal}, UERD~\cite{guo2015using}, \mbox{J-Mipod}~\cite{giboulot:hal-03096658},~...). 
Moreover, note that if the embedded payload is encrypted -- which is usually the case for security reasons -- decoding the embedded message is not possible as soon as one bit of the payload is changed.

\subsection{Prior Works on Robust Steganography}

\begin{table*}[h]
    \begin{centering}
    \begin{tabular}{|c|c|c|c|c|c|}
    \hline 
    Reference & Strategy & Errorless & Side information & ECC & Filtering before recompression \\
    \hline \hline
    \cite{kin2018adaptive} & Dual STC & No & No & Yes & No \\
    \hline
    \cite{zhang2016framework} & Watermarking & No & Yes & Yes & No \\
    \hline
    \cite{tao2018towards} and \cite{lu2020secure} & Coefficient adjustment & No & No & No & No\\
    \hline
    \cite{zhao2018improving} & Successive recompressions & No & No & Yes & No\\
    \hline
    \cite{ZENG22} & Non-robust set selection & No & No & Yes & Yes\\
    \hline
    \cite{WU22SSR} & Sign modification & No & Yes & No & No\\
    \hline
    Ours & Robust set selection & Yes & No & No & Yes\\
    \hline
    \end{tabular}
    \par\end{centering}
    \vspace{2mm}    
    \caption{Comparison between different schemes in the framework of steganography robust to JPEG compression.}
    \label{tab:comp}
\end{table*}

The domain of robust steganography aims at keeping the main constraint of steganography (\textit{i.e.} to embed an undetectable payload) but also adds the constraint of robustness, which can be defined as minimizing the bit error rate on the decoded payload after a lossy transmission channel. 
Note that this second constraint (robustness) is very similar to the one defined in watermarking. For this reason, secure watermarking~\cite{bas:hal-01283066} could hypothetically be considered as an option to embed robust and undetectable payload~\cite{gipsa:tifs07security}. 
However, the proposed schemes in the watermarking literature never considered either steganalysis to benchmark undetectability or large embedding rates. In watermarking, the payload characterizes an identifier of several dozens of bits, not a message of several kilobits.

In~\cite{kin2018adaptive}, Cleaves and Ker both study the impact of lossy transmission combined with syndrome trellis code (STC)~\cite{filler2011minimizing}. They show that the STC replicates the errors associated with the channel on the decoded payload. They propose to use a dual-STC scheme combined with Reed-Solomon (RS) Error Correcting Codes (ECC) to reduce the error rate while minimizing the embedding distortion. This scheme is, however, not benchmarked w.r.t. steganalysis. 

Zhang {\it et al.} proposed to mix a watermarking scheme~\cite{zhang2016framework} based on the modifications of DCT coefficients~\cite{luo2002fast} and a steganographic scheme~\cite{holub2014universal} to favor embedding on coefficients which are both robust and secure by weighting the cost related to J-UNIWARD. Unfortunately, the proposed scheme is very detectable (\textit{e.g.} $P_E=5\%$ at quality factor QF = 75 and 0.1~bpnzAC). The payload also needs to be protected using RS-ECC, and the error rate is still essential for large QF ($19\%$ at QF = 95).

In~\cite{qiao2021robust}, Qiao {\it et al.} propose to select robust cover elements, which are defined as robust because they are not equal to zero after double compression. Unfortunately, this scheme suffers from at least two drawbacks: 1) Alice has to transmit the set of robust elements to Bob as side information, and 2) the payload is still subject to errors, and the detectability is, by a large amount, more important than the non-robust scheme. 

Tao {\it et al.} propose the idea to generate an {\it intermediate image} after embedding by the mean of ``coefficient adjustment"~\cite{tao2018towards}. This image is a modified version of the stego image, and the modifications are computed to cancel the modifications due to an ideal JPEG coding scheme. A similar idea to invert the JPEG compression scheme was developed by Lu~\mbox{\it et al.} by combining coefficient adjustment with an auto-encoder predicting the input image~\cite{lu2020secure}. In both cases, the changes made to the intermediate image increase the detectability of the modified stego image. The practical implementation of finding a perfect intermediate image is also questioned in~\cite{lu2020secure} due to convergence issues. The authors adopt another strategy using the auto-encoder, but it cannot cancel all the modifications due to coding.

Recently, Zhao {\it et al.}~\cite{zhao2018improving}, proposed successive compressions of the cover image to reduce the number of changes after embedding and compression. This method is effective but has the disadvantage of generating (recompressed) cover images that are different from natural ones, hence more detectable. The proposed scheme also uses BCH-ECC to decrease the error rate after embedding.  

One of the most recent works, Sign Steganography Revisited (SSR)~\cite{WU22SSR}, uses the sign of the DCT coefficients to communicate the secret message. The method is checking for every DCT mode if there is a coefficient that will change its sign during recompression. If that is the case, it will prohibit embedding into this DCT mode. To this end, the steganographer needs to additionally communicate the robustness of all 64~DCT modes as a side-information. 

The last line of research on this topic is the scheme called MINImizing Channel Error Rate (MINICER)~\cite{ZENG22}. The algorithm first recompresses the cover images, then checks if an embedding change would create a so-called `overflow', which means that there is a pixel with value outside of the interval [0,255] after decompression. If so, the algorithm considers such embedding change as non-robust by setting its embedding cost to infinity. The recompressed cover image is then embedded with modified costs and specific DCT coefficients are changed back to the original single-compressed cover value in order to create a single-compressed stego image. This algorithm can also deal with filtering before recompression. 

While both SSR and MINICER, could potentially provide small bit error rates in some limited scenarios, we will show in Section~\ref{subsec:success} that in a practical setting, they are, in fact, not robust.

\begin{figure*}[h]
    \centering
    \includegraphics[width=\linewidth]{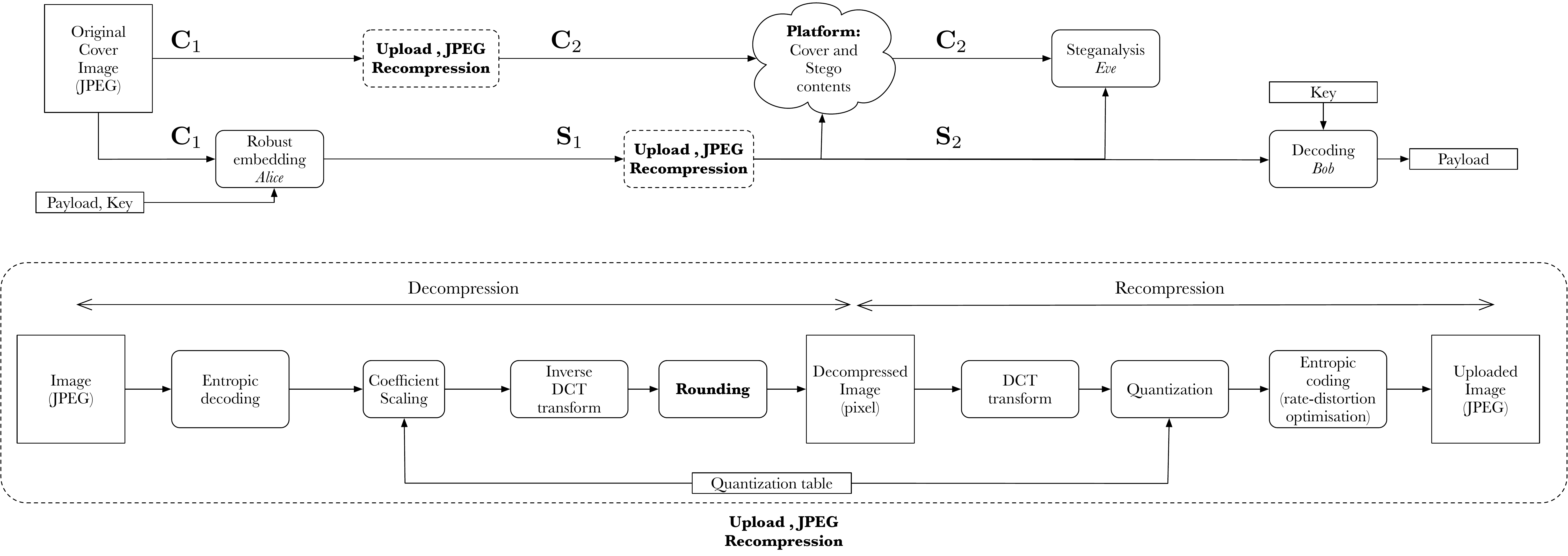}
    \caption{Considered setups for robust steganography after JPEG compression: the top diagram represents the whole chain of processes and the different players (Alice, Bob, and Eve), see also Section~\ref{subsec:security}. The bottom diagram depicts the different operations involved in the uploading process, see also Section~\ref{subsec:recompression}.}
    \label{fig:setting}
\end{figure*}

It is important to note that, due to the practical context of steganography, the scheme's robustness has to be extremely high, as only one erroneous bit can jeopardize the whole transmission. Errorless steganographic schemes are consequently recommended. Regarding robustness to image scaling, several works address this issue for different downsampling kernels, such as the nearest neighbor kernel or the bilinear kernel, where only pixels that are respectively preserved (see Zhang {\it et al.}~\cite{zhang2018robust}), or contribute the most (see Zhu {\it et al.}~\cite{zhu2021inverse}), are modified. To the best of our knowledge, errorless robust JPEG steganography has not been investigated before.   



\subsection{Outline of the Paper}

This paper proposes an errorless steganographic scheme in the JPEG domain robust to JPEG recompression. As summed up in table~\ref{tab:comp}, the advantages of this scheme are to be errorless (it can be guaranteed that generated stego will not produce any error at the embedding), so not convey any side-information for payload extraction, to not rely on the use of Error Correcting Codes (ECC). Moreover, the embedding enables to guarantee robustness even when filtering is applied before re-compression. 

Section~\ref{sec:setups} presents both the security setup (\textit{i.e.} the knowledge of the embedder) and the coding setup (\textit{i.e.} the JPEG coding process). Section~\ref{sec:extractRob} details an algorithm proposed to use the output of JPEG coders to extract a set of robust coefficients according to a specific scanning strategy. Section~\ref{sec:embAlgo} presents the embedding and decoding algorithms together with a strategy to spread the payload into 64 lattices. Section~\ref{sec:results} presents results on the detectability of the proposed scheme and compares it with naive embedding. A robustness analysis is also proposed when JPEG coding uses rate-optimization strategies. 


\section{Considered Setups}\label{sec:setups}

\subsection{Notations}\label{subsec:notations}
A bold capital letter is considered as a matrix, and a lowercase letter denotes a coefficient. $(i,j)$ denotes the pixels or coefficients coordinates $i$ and $j$.

\subsection{Security Setup}\label{subsec:security}

Robust steganographic schemes can be split into two categories, schemes which assume that the compression parameters are either known or unknown. The scheme that we present here belongs to the first category. 
Its security setup is illustrated on the top diagram of Fig.~\ref{fig:setting}.

Alice, the embedder, sends a stego image, denoted $\mathbf{S}_1$, on a platform that compresses $\mathbf{S}_1$ into $\mathbf{S}_2$ using a JPEG coder. Bob, the receiver, downloads the image from the platform and tries to decode the payload. Because we assume that Alice's cover, denoted $\mathbf{C}_1$, is also in the JPEG format, the image is consequently double-compressed. Note that the setup is equivalent to classical (lossless) steganography if the platform does not recompress the uploaded image. In the following, we consider that the platform does compress the uploaded image.

We assume that Alice knows both the coding scheme and the coding parameters used by the platform. Practically, this can be done by inspecting the uploaded-downloaded images. The JPEG quantization matrix is public, and the coding scheme can often be identified by comparing the uploaded/downloaded image with the outputs of different coders.

Note that this setup follows the Kerckhoffs' principle, which states that anything not related to the secrecy of the application (here, the fact that a payload is potentially transmitted to Bob) should be considered public. 
Within this setup, the steganalyst Eve has access to the platform. She can also download images to try to differentiate an uploaded cover content (denoted $\mathbf{C}_2$) from an uploaded stego content ($\mathbf{S}_2$). 

In order to minimize the difference between the uploaded image (respectively $\mathbf{C}_1$ or $\mathbf{S}_1$) and the downloaded image (respectively $\mathbf{C}_2$ or $\mathbf{S}_2$), we assume that the uploaded image is already coded with the same coding parameters than the downloaded image. We assume first that JPEG compression is the only process done during the upload on the platform, which means that the uploaded image is already resized correctly in order to prevent any resizing operation. This assumption is rather practical since it is usually the case on many platforms such as Facebook, WhatsApp, FlickR~\cite{castiglione2011forensic,caldelli2017image} or {\it Slack}. Note that in section~\ref{subsec:filtering} we also consider the case where a size-invariant filter is applied to the image before compression. 

\subsection{JPEG Coding Setup}\label{subsec:recompression}


\begin{figure}[t]
    \centering
    \includegraphics[width=\columnwidth]{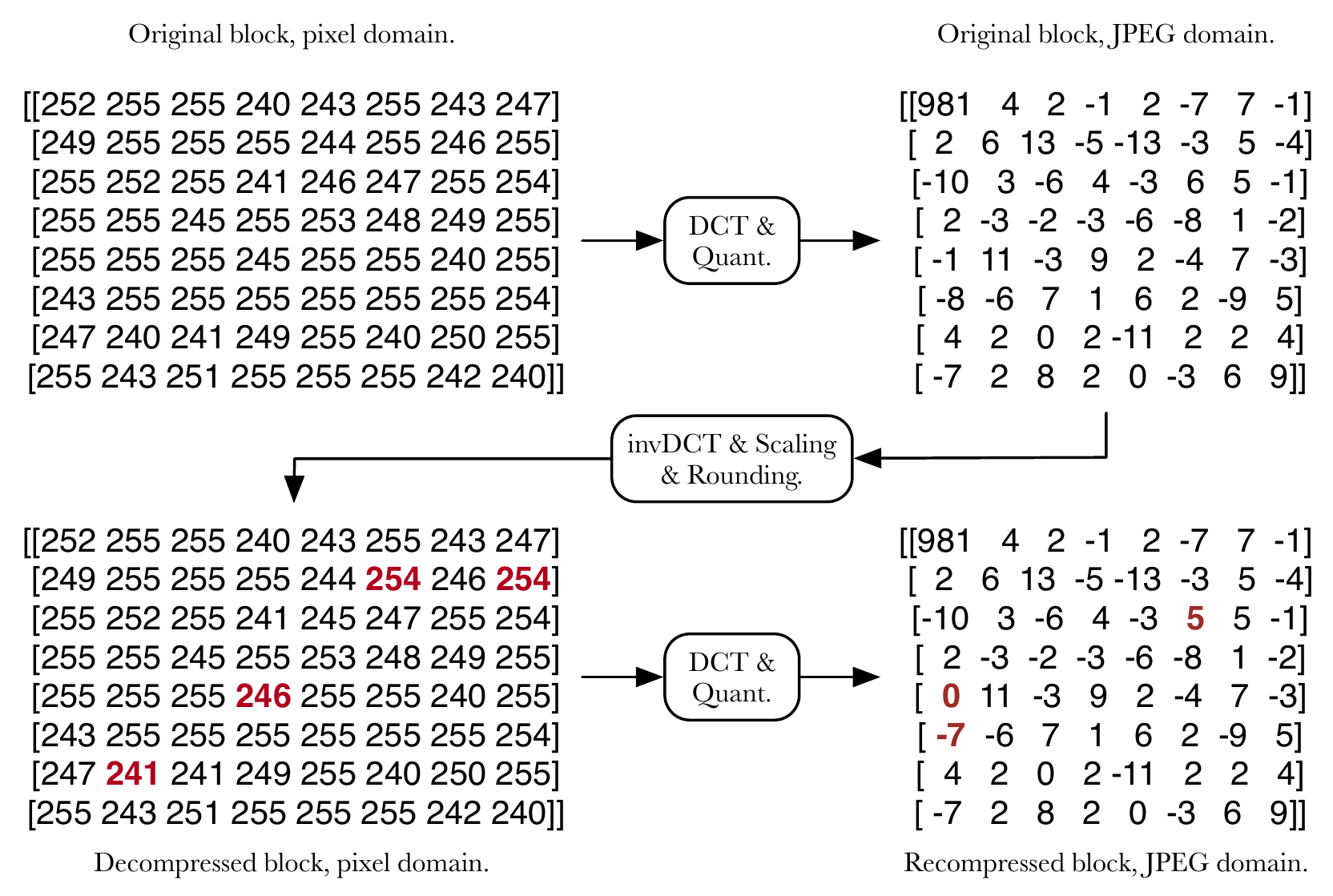}
    \caption{Example of DCT coefficient changes after decompression and recompression (QF = 100). Changes in the pixel or JPEG domain are displayed in bold and dark red.}
    \label{fig:clippingProblem}
\end{figure}

Firstly, we recall the main features of the JPEG coding scheme. Without loss of generality, we assume that the original image is coded as greyscale. Still, the same methodology regarding the embedding mechanisms described in Sections~\ref{sec:extractRob} and~\ref{sec:embAlgo} can be applied on color channels, with or without sub-sampling. 

- The original image is first decomposed into disjoint blocks of size $8 \times 8$ pixels. 

- Each block is then transformed into 64 DCT coefficients using the specific DCT type-II transform. 

- Each coefficient is then quantized according to a quantization matrix specific to the coding algorithm and each DCT mode. 

Note that there is a specific relation between the JPEG quality factor QF and the quantization matrix, but depending on the implementation of the coder, this relation can be different. For example, the Libjpeg library\footnote{\url{http://libjpeg.sourceforge.net}}, associated with the \texttt{convert} command, uses the classical relationship proposed by the standard (see~\cite{butora2021revisiting}, Section~IV). On the contrary, the \texttt{mozjpeg} library\footnote{\url{https://github.com/mozilla/mozjpeg}} uses {\it ad hoc} quantization tables.

- For each block, the coefficients are scanned using a zigzag order.

- Depending on the coding scheme, the lossless entropic coding scheme can be different: the quantized coefficients are either directly coded using Run Length Coding and Huffman Coding, or a rate-distortion optimization procedure is applied to change the magnitude of several coefficients to increase the coding rate.       

Note that in the first case, the coefficients and blocks are independently coded, but in the second case, there is an interplay between the possible coefficient values and the length of the produced code. The potential use of the rate-distortion procedure depends on the implementation. For example, the Libjpeg library does not implement by default any rate-distortion procedure, but the \texttt{mozjpeg} library implements by default a Viterbi algorithm relying on a trellis. 

The upload of a JPEG image on the platform consists in decompressing and then recompressing the image (see Fig.~\ref{fig:setting}, bottom). Before recompression, the JPEG image is decoded by first performing entropic decoding, coefficient scaling according to the quantization matrix, and inverse DCT transform. This is a lossy process because once decompressed, the pixel values are rounded to integer values between 0 and 255. This rounding operation in the pixel domain can modify a fraction of the DCT coefficients after recompression. Such an effect will be particularly significant (but not only) on blocks containing pixels initially clipped to 255 since, after decompression, the clipping may change the magnitude of DCT coefficients (see example depicted in Fig.~\ref{fig:clippingProblem}).

\subsection{An example: {\it Slack} instant messaging}\label{subsec:slack}

In this section, we want to give a real-world example of where robust steganography is needed. As a representative, we chose the {\it Slack} application for instant messaging\footnote{\url{https://slack.com}}. During our experiments tested on MacOS Version 4.31.156, we learned that given a JPEG image, {\it Slack} does, in fact, recompress the image before transmitting it to the receiver. As such, standard steganographic tools cannot be used, as the recompression would destroy the embedded message. This paper aims to propose a steganographic scheme that is robust to such a process.  
We learned that the recompression in {\it Slack} produces the exact same DCT coefficients and quantization tables as if produced by \texttt{convert} (see Section~\ref{sec:embAlgo}) without specifying any compression quality parameters. 
\begin{center}
    \texttt{convert input\_name.jpg output\_name.jpg}    
\end{center}

This process is equivalent to the use of the closest standard quantization table of \texttt{input\_name.jpg} during the generation of \texttt{output\_name.jpg}. This allows us to simulate the {\it Slack} recompression by simply locally calling the \texttt{convert} compressor. All the results reported for \texttt{convert} are thus directly applicable to the communication on the {\it Slack} social network.


\section{Robust Embedding}\label{sec:extractRob}


\subsection{Overview of the Embedding Algorithm}\label{subsec:embAlgo}

\begin{figure*}[t]
    \centering
    \includegraphics[width=0.7\textwidth]{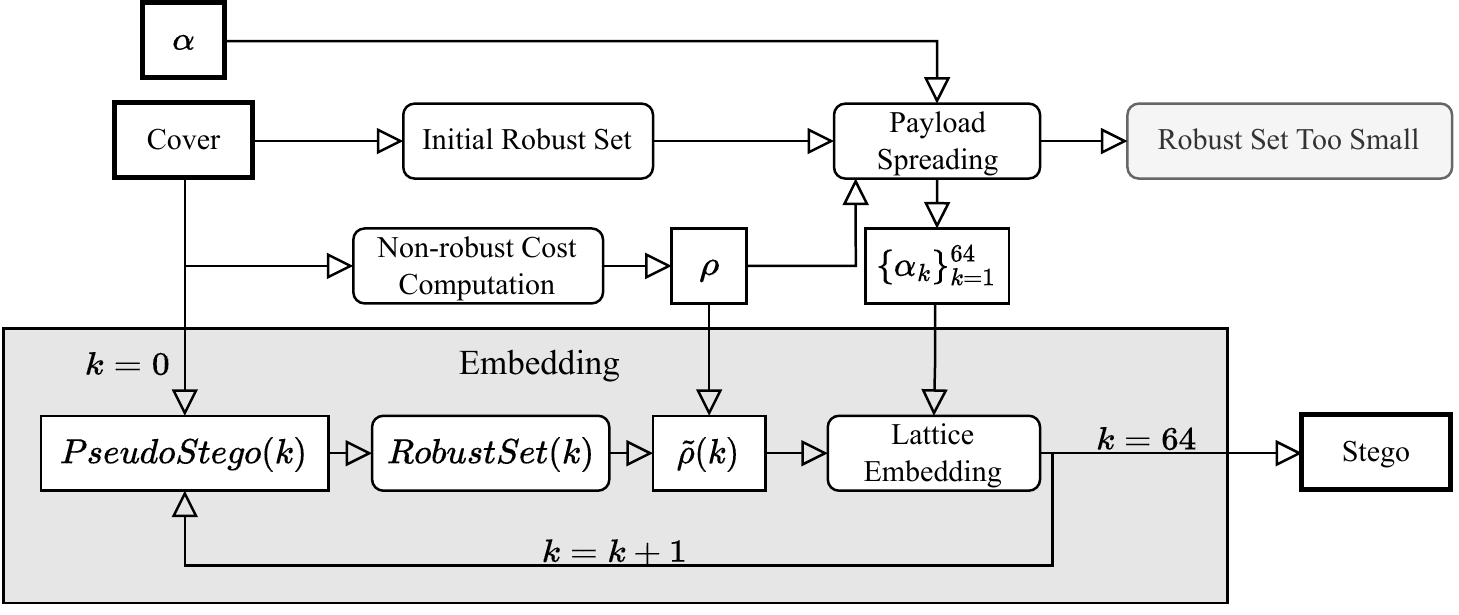}
    \caption{Main steps of the embedding algorithm.}
    \label{fig:stepEmbedding}
\end{figure*}

This section presents the robust embedding strategy. For better readability, we outline the general idea of the embedding scheme and only then describe the details of every mechanism involved.  

First, we define a robust coefficient and how to extract the set of robust coefficients. Since one embedding change can potentially affect the robustness of many DCT coefficients from the same $8\times8$ block, we divide the image into 64~non-overlapping lattices (one per DCT mode) and perform the embedding iteratively on every lattice separately. We assume for simplicity that every $8\times 8$ block is coded independently during the JPEG compression (we shall see in Section~\ref{sec:results} the impact of this assumption when it is not the case). 

Then, we show how to use the robustness of the coefficients during embedding. And finally, we look at how much payload shall be embedded in each of the 64 lattices.

The scheme of the embedding procedure is shown in Fig.~\ref{fig:stepEmbedding}.

\subsection{Robustness}

For ease of understanding, we introduce several definitions.

\begin{definition}[Processed modes]
Given $k$-th DCT mode (with a pre-defined ordering of modes), $k\in\{1,\ldots,64\}$, denote $\mathcal{P}_{k}=\{1,\ldots,k-1\},\,\mathcal{P}_{1}=\emptyset$ the set of all modes that have already been processed by the algorithm.
\end{definition}

\begin{definition}[Pseudo-stego]
The $k$-th pseudo-stego is the cover image with already embedded lattices from $\mathcal{P}_{k}$.  
\end{definition}

We will describe the process for a single $8\times8$ block of single-compressed DCT coefficients of the $k$-th pseudo-stego $\mathbf{c}=\{c_{n}\}_{n=1}^{64}\in\mathbb{Z}^{64}$. Let $i_{n}\in\mathbb{Z}^{64}$ be a vector containing \mbox{$i\in\mathbb{Z}$} at $n$-th coordinate and zeros elsewhere. Let \mbox{${R}(\mathbf{c},i_{n})\in\mathbb{Z}^{64}$} denote the recompressed DCT coefficients of $(\mathbf{c}+i_{n})$.\\ 

\begin{definition}[Robust coefficient]
We say a coefficient $c_{k}$ is robust towards an embedding change $i\in\{-1,1\}$, if during recompression it:\\

\begin{itemize}
	\item[] (R1): Does not change processed modes: $$\forall l\in\mathcal{P}_{k}:\,{R}(\mathbf{c},i_{k})_{l}={R}(\mathbf{c},\mathbf{0})_{l}.$$
	\item[] (R2): Preserves a change by $i$: $${R}(\mathbf{c},i_{k})_{k}=c_{k}+i.$$
	\item[] (R3): Preserves no change: $${R}(\mathbf{c},\mathbf{0})_{k}=c_{k}.$$
\end{itemize}
The sets of all robust coefficients towards $+1$ and $-1$ are denoted $\mathcal{R}_{k}^{+}$ and $\mathcal{R}_{k}^{-}$ respectively.

If a coefficient does not belong to a robust set, we say it is non-robust. The set of all non-robust coefficients is denoted~$\mathcal{R}_{k}^{0}$.
\end{definition}

Note that without (R1), embedding in the $k$-th lattice would destroy the message encoded in a lattice $l\in\mathcal{P}_{k}$, which would make the secret message unreadable. 

(R2) states that the embedding change needs to survive the recompression. 

(R3) gives us a choice during the embedding, whether to change the coefficient or not. We want to point out that a coefficient can belong to both sets $R_{k}^{+}$ and $R_{k}^{-}$. 

\begin{figure}[h]
    \centering
    \includegraphics[width=1\columnwidth]{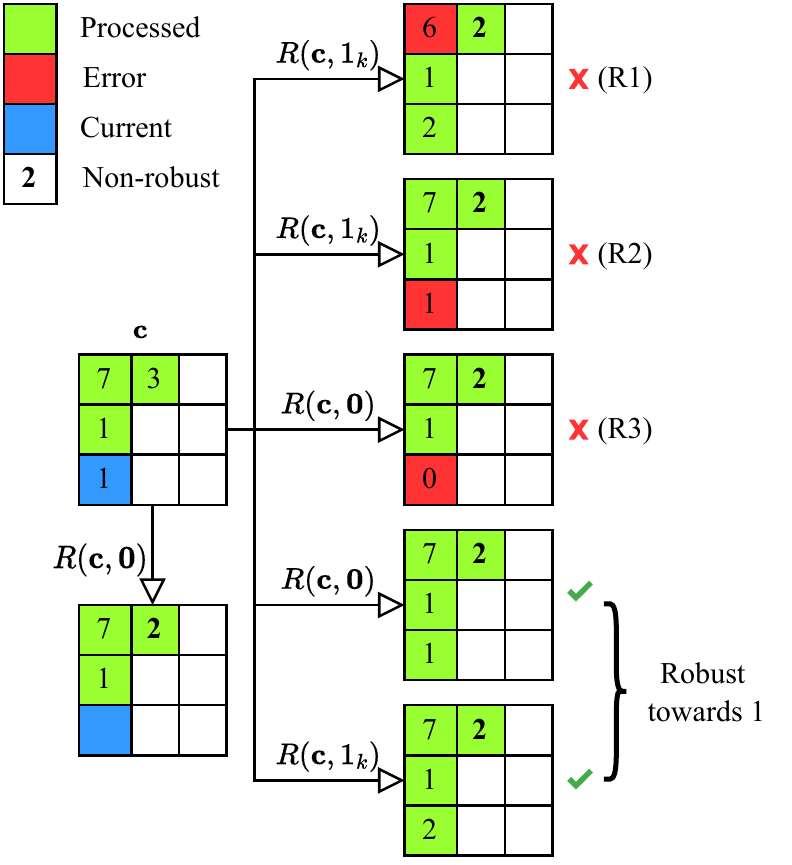}
    \caption{Mechanism for deciding robustness of a DCT coefficient. Left: DCT coefficients $\mathbf{c}$ and their recompressed values, Right: Possible effects of recompression on the current and processed lattices. The first three situations, each associated with one condition R1, R2, or R3, assign the coefficient to a non-robust set. Only if the last two situations arise is the coefficient robust towards 1.}
    \label{fig:stepRobustCoeff}
\end{figure}

We illustrate these conditions graphically with a $3\times3$ DCT block (for the sake of simplicity) in Fig.~\ref{fig:stepRobustCoeff}. 

From the construction of the robust sets, the non-robust coefficients are coefficients $c_{k}$, such that either ${R}(\mathbf{c},\mathbf{0})_{k}\neq c_{k}$, or ${R}(\mathbf{c},i_{k})_{k}\neq c_{k}+i,\,i\in\{-1,1\}$.  In either case, it is essential to note that even though we cannot have control over the recompressed coefficient, we can compute its value ${R}(\mathbf{c},\mathbf{0})_{k}$. We can see that this was, in fact, the case for one of the already processed modes (0,1) in Fig.~\ref{fig:stepRobustCoeff}: its value changed from 3 to 2, but it does not prevent us from correctly embedding. For practical implementation with STCs, we perform standard embedding on the robust sets (with their possible embedding values) and do not change the non-robust set. This is done by setting their corresponding embedding costs to wet costs. Since we do have access to ${R}(\mathbf{c},\mathbf{0})$, we are still able to encode the message in the trellis. Section~\ref{sec:embAlgo} explains the embedding mechanism in further detail.

\subsection{Lattice Embedding}\label{subsec:latEmb}

To embed a given lattice, we need to compute the robust costs. Let $\rho^{+},\rho^{-}$ denote the embedding costs (computed from a given steganographic algorithm) of changing a DCT coefficient $c$ by $+1$ or $-1$. The robust costs $\tilde{\rho}^{+},\tilde{\rho}^{-}$ are created by updating the original costs:

\begin{equation}
\begin{cases}
\tilde{\rho}^{\pm}=\rho^{\pm}, & c\in\mathcal{R}^{+}\cap\mathcal{R}^{-},\\
\tilde{\rho}^{-}=\infty, & c\in\mathcal{R}^{+},\\
\tilde{\rho}^{+}=\infty, & c\in\mathcal{R}^{-},\\
\tilde{\rho}^{\pm}=\infty, & c\in\mathcal{R}^{0}.
\end{cases}
\label{eq:robust_cost}
\end{equation}

In other cases, we keep $\tilde{\rho}^{+}=\rho^{+}$ and $\tilde{\rho}^{-}=\rho^{-}$.

Let $\alpha_{k}$ be the portion of the total payload we desire to embed in the $k$-th lattice. For practical embedding, we would provide the embedding costs and payload to STCs to perform the embedding. However, in this work, we simulate the optimal embedding. Therefore we find the optimal change rates:

\begin{equation}
\beta_{k}^{\pm} = \frac{e^{-\lambda\tilde{\rho}_{k}^{\pm}}}{1+e^{-\lambda\tilde{\rho}_{k}^{+}}e^{-\lambda\tilde{\rho}_{k}^{-}}},
\label{eq:optimal_betas}
\end{equation}

where $\lambda>0$ is the Lagrange multiplier ensuring that we embed the desired payload

$$\sum_{n=1}^{N}H_{3}(\beta_{n}^{+},\beta_{n}^{-})=\alpha_{k},$$

and $H_{3}(\beta^{+},\beta^{-})$ is the ternary entropy function
\begin{multline*}
H_{3}(\beta^{+},\beta^{-})=\\
-(1-\beta^{+}-\beta^{-})\log(1-\beta^{+}-\beta^{-})-\beta^{+}\log\beta^{+}-\beta^{-}\log\beta^{-}.
\end{multline*}
 Having the optimal change rates, the coefficients from this lattice are embedded. That concludes the embedding of the given lattice, and we move on to the next.

\subsection{Payload Spreading}\label{subsec:payAlloc}

Since our proposed method embeds iteratively 64 non-overlapping lattices, we need to decide \textit{a priori} what portion of the embedding payload is carried in every lattice. Therefore, we compute the so-called ``Initial robust set" for every lattice -- the robust set without any embedding change. Let $\alpha$ denote the total payload in bits we want to embed. Having the robust sets $\mathcal{R}^{+},\mathcal{R}^{-},\mathcal{R}^{0}$, we update the embedding costs (Eq.~\eqref{eq:robust_cost}) and compute the optimal change rates (Eq.~\eqref{eq:optimal_betas}) for the whole image. From these change rates, we compute the proportion of payload $\alpha_{k}$ in every lattice: $$\alpha_{k}=\sum_{n=1}^{N}H_{3}(\beta_{n,k}^{+},\beta_{n,k}^{-}),$$

where $\beta_{n,k}^{\pm}$ is the $n$-th change rate in the $k$-th lattice. At this stage, it is possible that we cannot communicate the desired payload because of the small size of the robust set. In such a situation, we have no choice but to use a different image or a smaller secret message. 

We are also well aware that there could be another potential issue during actual embedding with this approach of spreading the payload among lattices. In particular, the robust set in the $k$-th lattice could be of a very different size after embedding the previous lattices. As a result, we would not have enough robust coefficients to use for embedding a prescribed payload $\alpha_{k}$. However, in practice, we observed that the size of robust sets in each lattice changes in a negligible way, see Fig.~\ref{fig:rob_vs_lattice}.


\subsection{Filtering before recompression}\label{subsec:filtering}
The presented algorithm can be easily extended to be robust to a filtering operation (\textit{e.g.} blurring, sharpening, ...) occurring before the re-compression. This processing can be done on the platform to improve the rendering of the image once published. 

\begin{figure}[h]
    \centering
    \includegraphics[width=0.5\columnwidth]{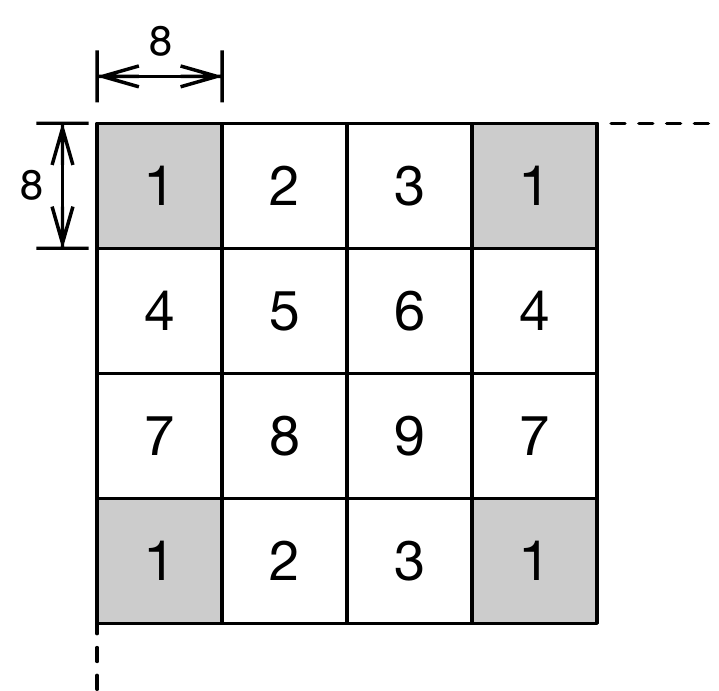}
    \caption{The 9 macro-lattices used to be robust to filtering.}
    \label{fig:Ninelattices}
\end{figure}

Assuming that the filter window size is smaller or equal to $8\times8$ pixels, this filtering operation can propagate changes between one block and its 8 neighbors, but not between non-adjacent blocks. We need in this case to embed separately in 9 extra macro-lattices, \textit{i.e.} sets of JPEG blocks that are distant by 16 pixels in one or two directions as depicted in Fig.~\ref{fig:Ninelattices}. The final number of lattices in this case is consequently multiplied by 9. 

One might wonder why 4 macro-lattices are not enough. The answer relies on the fact that each mode is affected by the same $\{-1,0,+1\}$ change and recompression. Consequently, if after one embedding operation, one change occurs in a block belonging to a previously visited macro-lattice, there is an ambiguity to know which of the two neighboring blocks is responsible for the change.  

\subsection{Number of calls to the compressor}\label{subsec:number_of_call}

In the end, with only recompression the embedding requires $3\times 64\times2$ calls to the compressor : one for each embedding modification in the set $\{-1,0,1\}$ multiplied by the 64 DCT modes multiplied by the number of steps necessary to perform the embedding, \textit{i.e.} the estimation of the robust set for payload spreading and the payload embedding.   
If the robustness has to deal with filtering and compression, then the number of calls equals $9\times3\times64\times2$. Note that a call to a JPEG compressor is rather fast and the (UERD) embedding operation robust to the compressor takes about $10$s on a Macbook Pro with M1 chip on a $512\times512$ image and the \texttt{convert} compressor, and $90$s when filtering is involved. Note that the filtering makes the scheme 9 times slower because we need to consider 9 times more lattices, as explained in the previous section.

\section{Practical Implementation}\label{sec:embAlgo}

\begin{figure}[h]
    \centering
    \includegraphics[width=0.8\columnwidth]{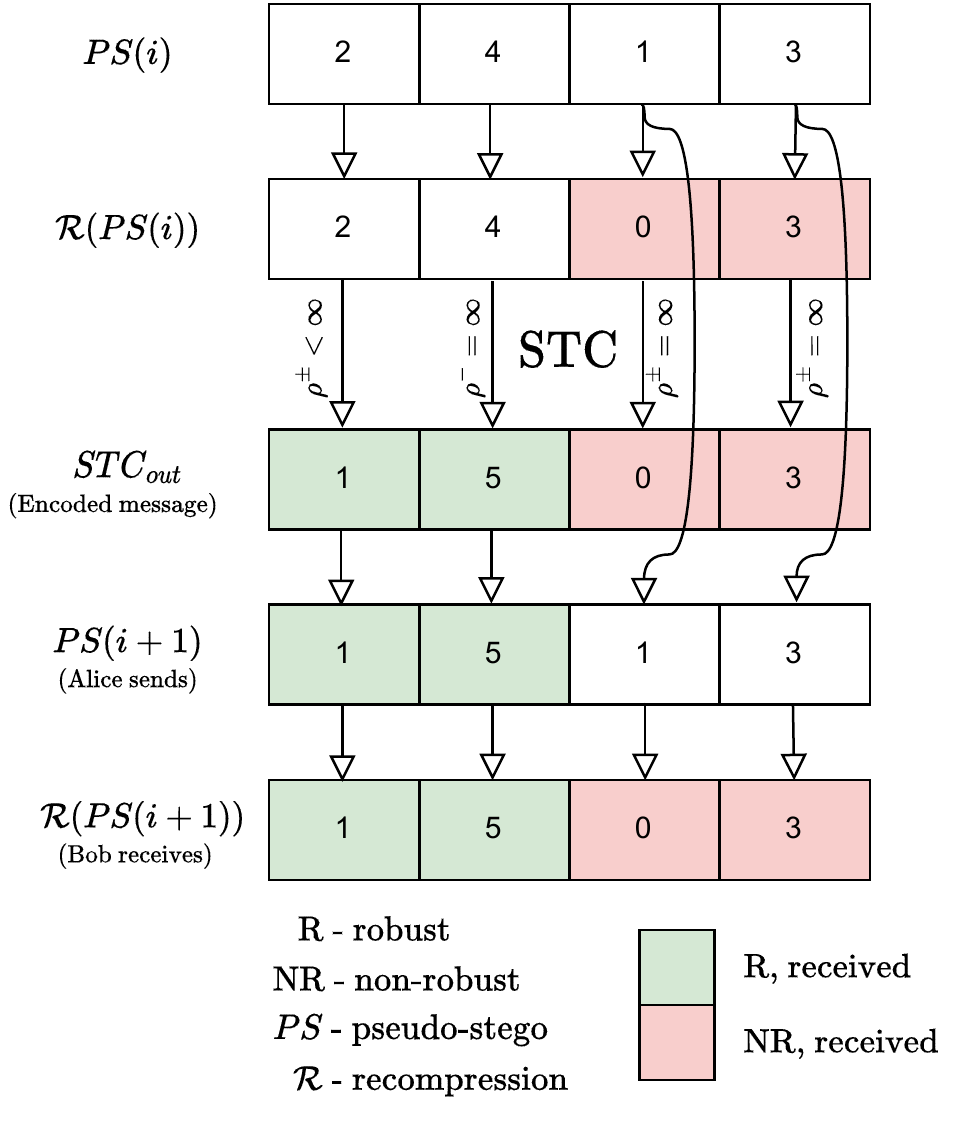}
    \caption{Practical embedding of $i$-th lattice with Syndrom-Trellis Codes: Alice sends a bitstream which takes into account the recompression $\mathcal{R}(.)$ in order to be compliant with both the STC coding and decoding processes. The costs $\rho$ are also chosen in order to deal with non-robust modifications.}
    \label{fig:trellis}
\end{figure}

\begin{figure}[h]
    \centering
    \includegraphics[width=0.8\columnwidth]{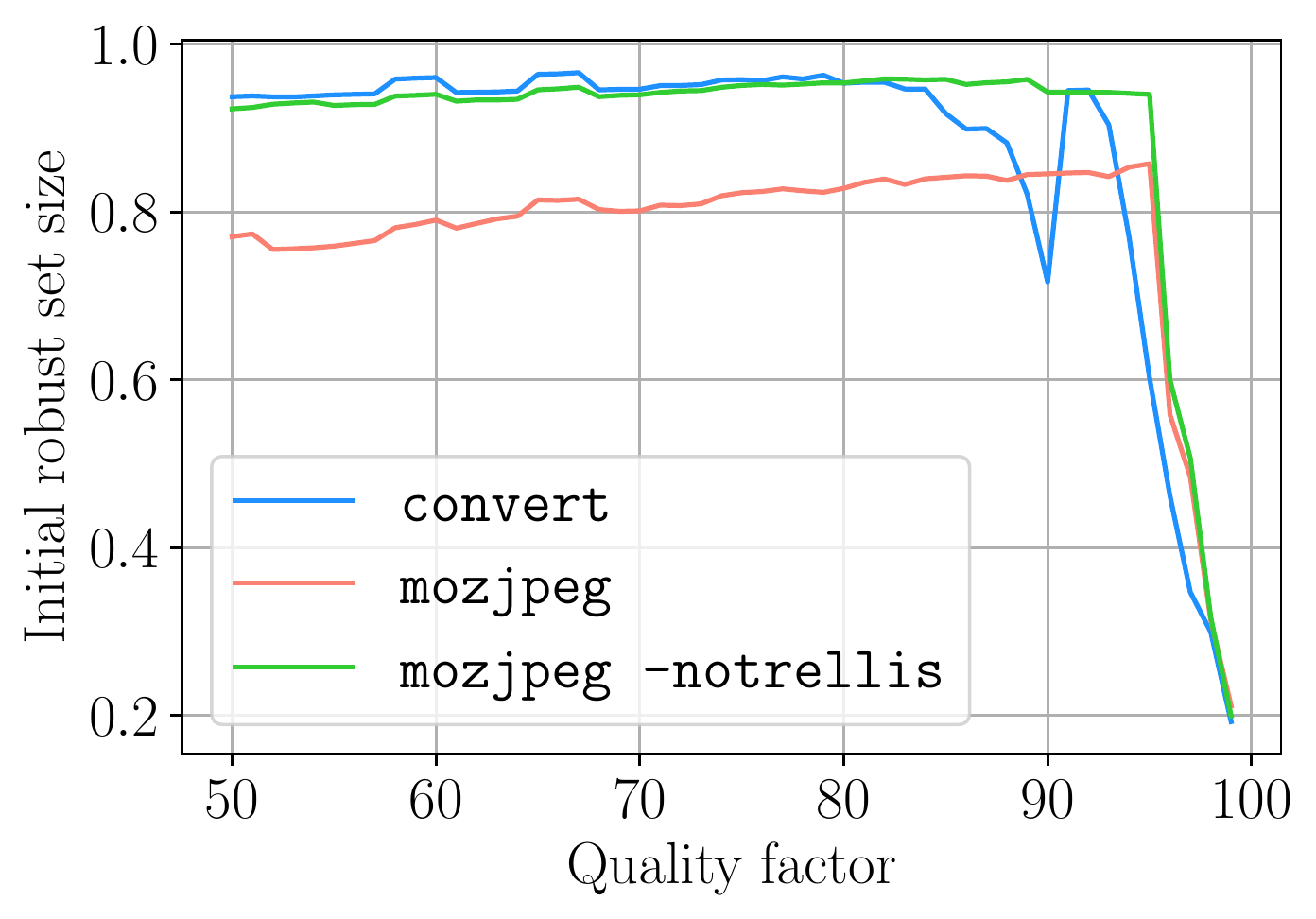}
    
    \includegraphics[width=0.8\columnwidth]{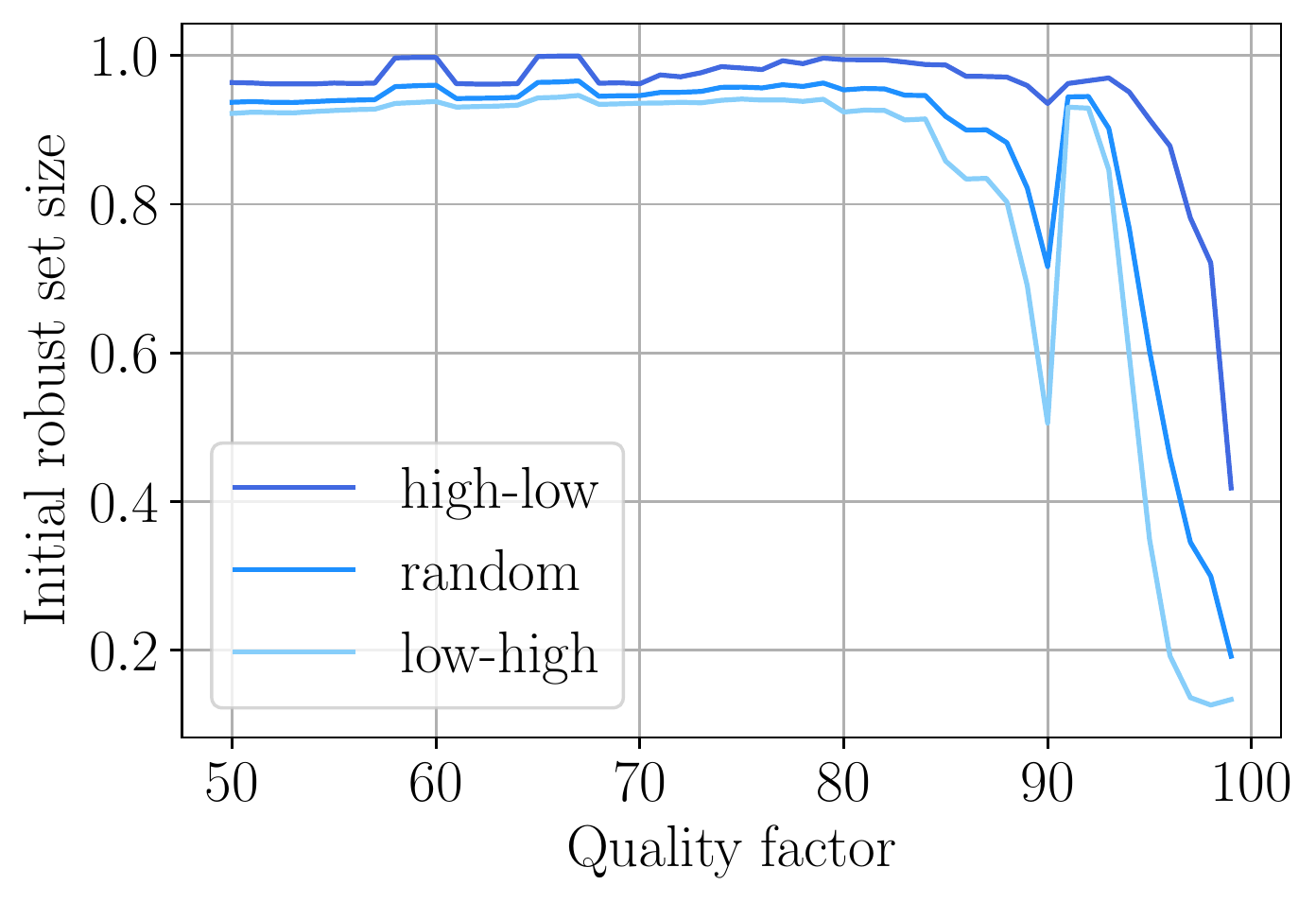}
    \caption{Relative robust set size average over 10 cover images. Top: Random scanning strategy, bottom: \texttt{convert}.}
    \label{fig:robust_size}
\end{figure}

\begin{figure}[h]
    \centering
    \includegraphics[width=0.5\columnwidth]{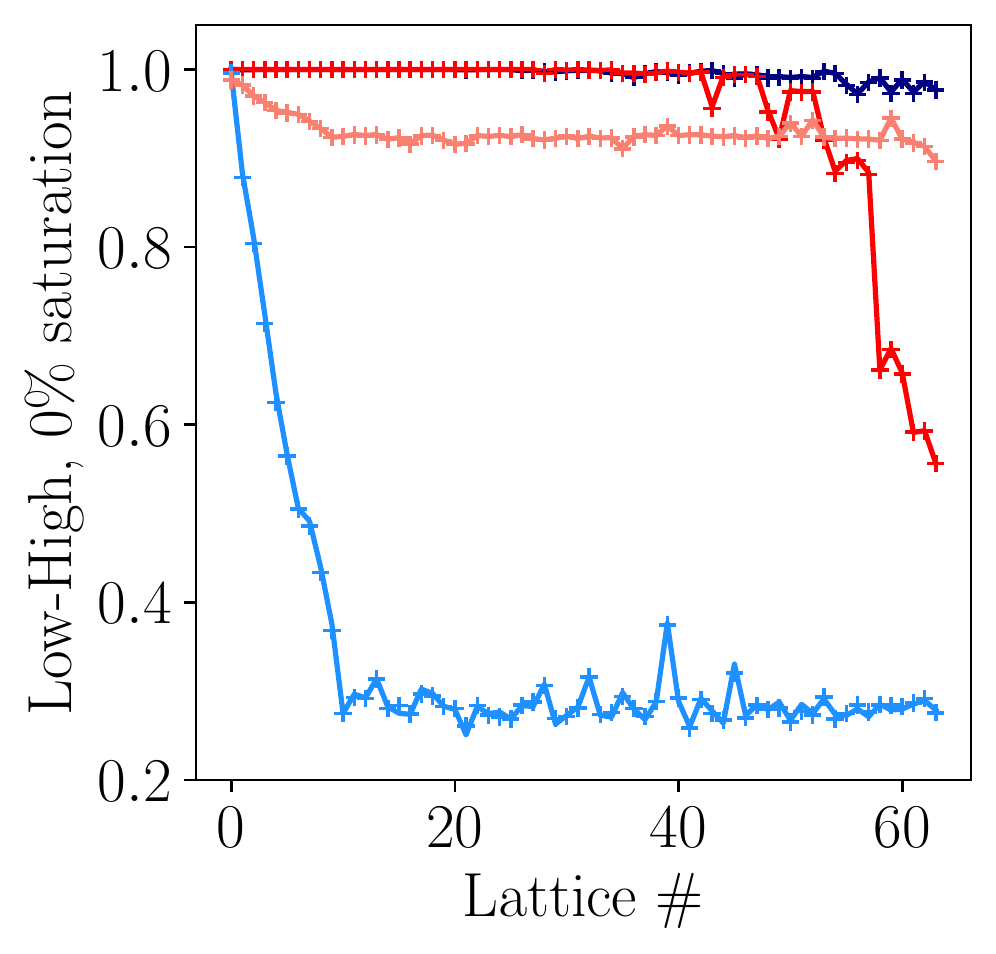}\includegraphics[width=0.5\columnwidth]{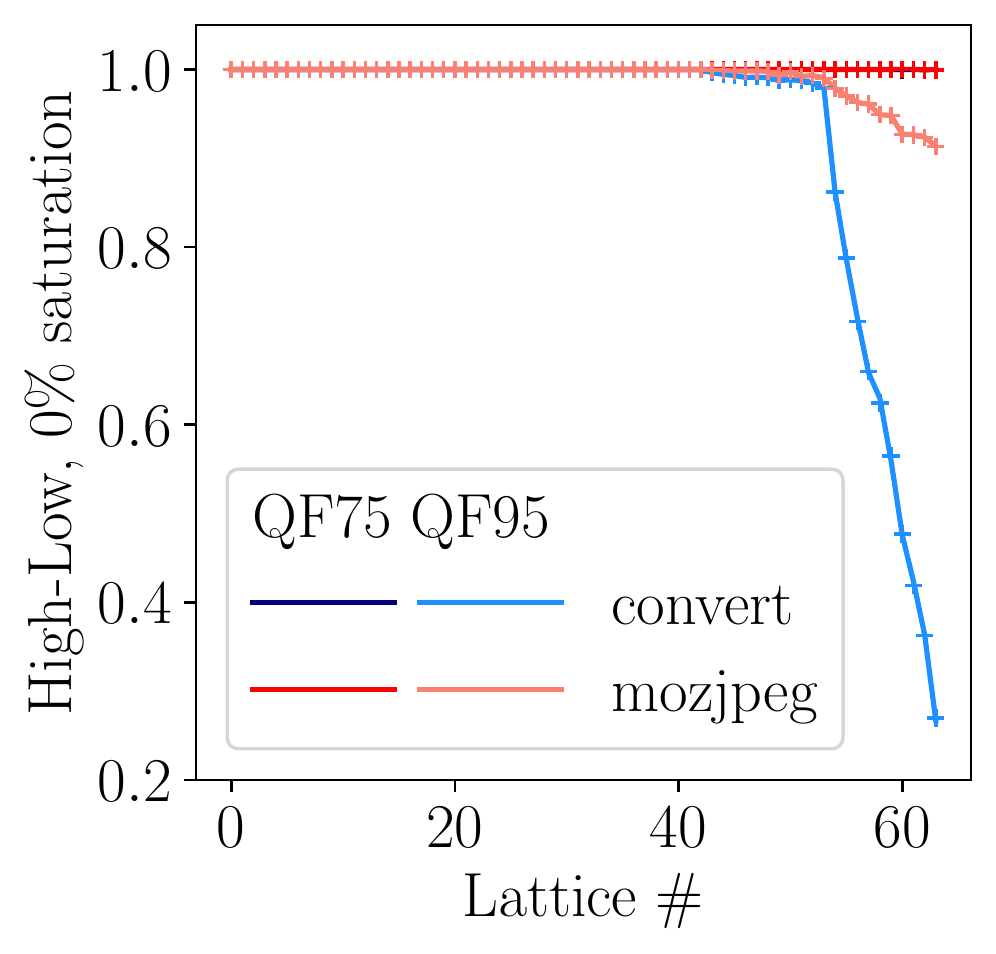} \\
    \includegraphics[width=0.5\columnwidth]{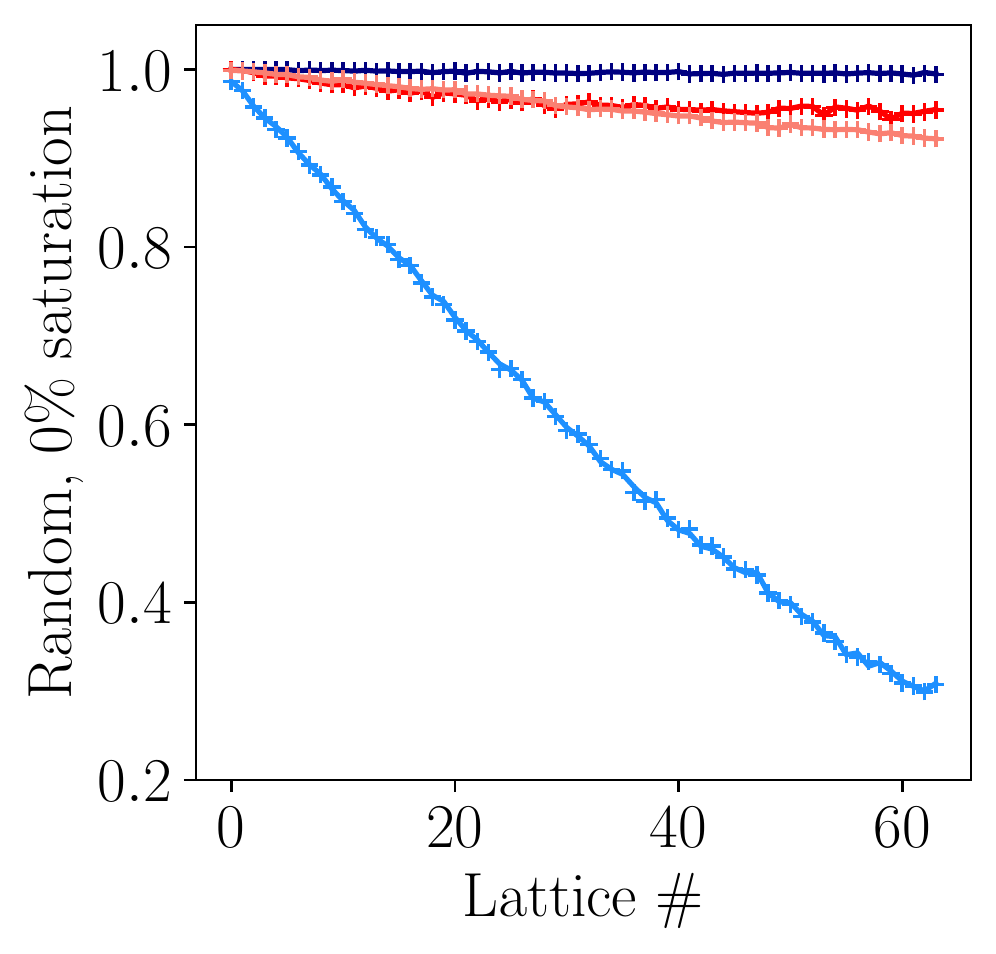}\includegraphics[width=0.5\columnwidth]{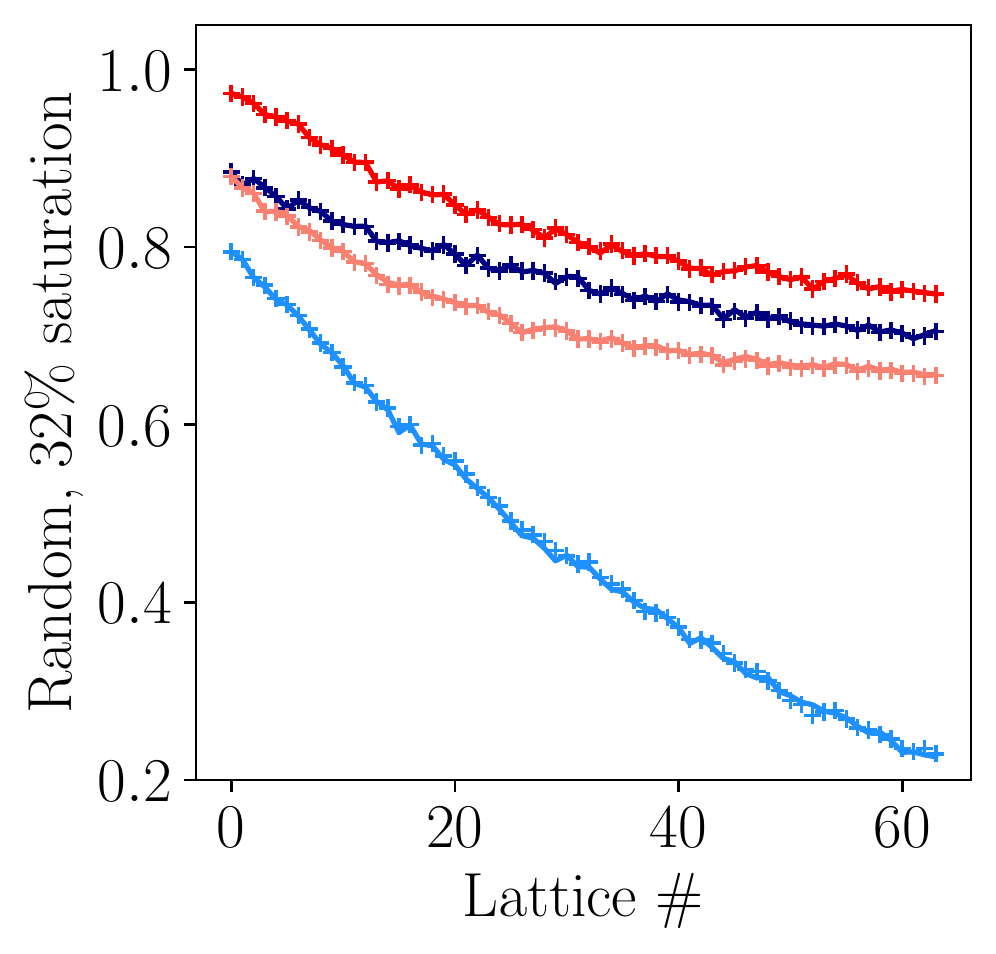}
    \caption{Proportion of robust coefficients in every lattice during embedding. The solid line corresponds to a cover image (Initial robust set), while the crosses mark a stego image embedded with 0.8 bpnzAC. The first three figures are from a single BOSSBase image of size $512\times512$ without any saturated pixels across three scanning strategies. The last figure shows the robustness of an image with 32\% of its pixels saturated and random scan.}
    \label{fig:rob_vs_lattice}
\end{figure}

\begin{figure*}[h]
    \begin{centering}
    \includegraphics[width=1.0\textwidth]{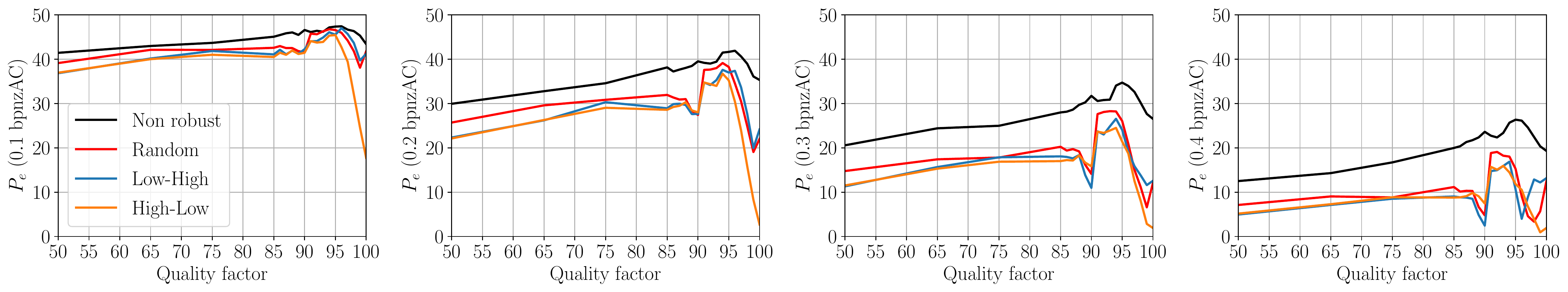}
    \caption{Detection errors with \texttt{convert}.\label{fig:pract_sec_convert}}
    \par\end{centering}
\end{figure*}

\begin{figure*}[h]
    \begin{centering}
    \includegraphics[width=1.0\textwidth]{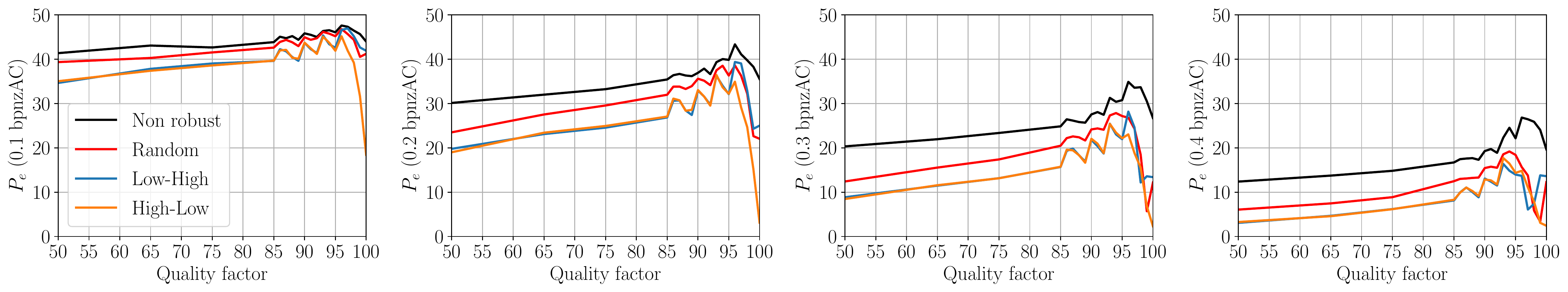}
    \caption{Detection errors with \texttt{mozjpeg} without rate-distortion optimization.\label{fig:pract_sec_mozjpegnotreillis}}
    \par\end{centering}
\end{figure*}

\begin{figure*}[h]
    \begin{centering}
    \includegraphics[width=1.0\textwidth]{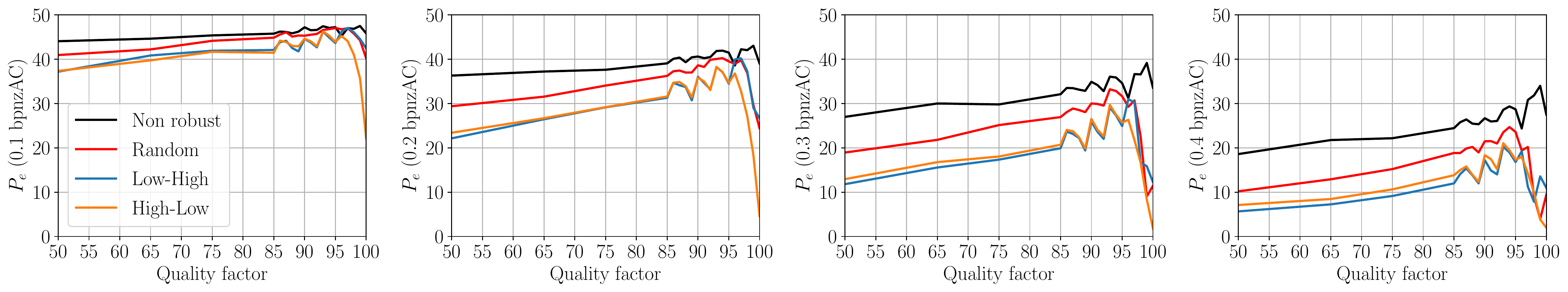}
    \caption{Detection errors with \texttt{mozjpeg} with rate-distortion optimization.\label{fig:pract_sec_mozjpegtreillis}}
    \par\end{centering}
\end{figure*}

In this section, we want to detail the technicalities necessary for actual embedding. First, JPEG compression is significantly impacted by the compressor. We consider three different JPEG compressors in this work: ImageMagick's \texttt{convert}, \texttt{mozjpeg} (by default with rate-distortion optimization), and \texttt{mozjpeg} without optimization ( \texttt{-notrellis} option). As mentioned earlier in Section~\ref{subsec:recompression}, the main difference between these compressors is the quantization tables used. Specifically, \texttt{mozjpeg} uses stronger quantization (bigger quantization steps) than \texttt{convert}. This harsher quantization positively affects the robustness of an image. In Fig.~\ref{fig:robust_size}, we show the average initial robust size over 10 images compressed with the three compressors across a range of quality factors and with the random scanning strategy (see Section~\ref{subsec:scan} for more details about scanning strategies). We can see that for quality factors above $80$, the images compressed with \texttt{convert} have a smaller robust set than those compressed with \texttt{mozjpeg} without the rate-distortion optimization. This is especially true around quality factor $90$, where \texttt{convert} has a sudden drop in the robust set size. Strangely enough, \texttt{convert}'s robust size jumps back $93\%$ for qualities $91$ and $92$. Although we are not sure why exactly this phenomenon is happening, we are convinced it is related to the quantization tables because we are unaware of another substantial difference between the compressors. Additionally, we can notice that disabling \texttt{mozjpeg}'s rate-distortion optimization increases the robust set size, mainly for qualities below $95$. 

\subsection{Scanning Strategies}\label{subsec:scan}

Secondly, we can notice that in the definition of {\it Processed modes}, we assumed a given ordering of DCT modes. We consider three natural scanning strategies: 
\begin{enumerate}
\item Low-High: Scan modes in a zig-zag manner from low to high-frequency modes (as done in JPEG),
\item High-Low: Reverse the order of Low-High, and
\item Random: Randomly assign ordering of modes in every $8\times 8$ block.\footnote{The pseudorandom key used for generating the permutations can be a part of the secret key.} 
\end{enumerate}

We will see in Section~\ref{sec:results} that these three strategies will affect the empirical security of the embedding scheme. 

Next, a scanning strategy dictates the size of the robust set in every lattice. In Fig.~\ref{fig:rob_vs_lattice}, we show the relative size of the robust set across all 64 lattices. The top two plots show the Low-High and High-Low scanning strategies for an image without saturated pixels, while the bottom plots show the Random scanning strategy for a non-saturated and a greatly saturated image. We can make several interesting observations from this figure. Robustness at QF 95 is generally smaller than for QF 75. This is to be expected, as bigger quantization steps (lower quality) provide more robustness against recompression. Similarly, since \texttt{mozjpeg}'s quantization steps are mostly bigger than those of \texttt{convert}, \texttt{mozjpeg} yields on average better robustness. We observe that the Random scan linearly decreases the robustness as we proceed through the lattices. It is important to realize that in the Random strategy, one lattice does not correspond to a DCT mode, as every block is scanned in a different random order. This ensures a relatively equal payload across all lattices, which allows us to skip the computation of the Initial robust set. In contrast, the first two scanning strategies assign one lattice per DCT mode, and it can be seen (especially for QF 95, \texttt{convert}) that most of the payload is carried in the low-frequency modes because their robustness decreases drastically. Finally, the saturation of pixels reduces the robustness due to the clipping of pixels into the dynamic range $[0,255]$ during recompression. During our experiments, all non-saturated images exhibit a similar trend, while robustness tends to decrease with increasing saturated area.
Interestingly, we can also notice that the size of the Initial robust set in every lattice is not very different from the robust set of an image embedded with $0.8$ bpnzAC. This justifies our payload allocation based on the Initial robust set (see Section~\ref{subsec:payAlloc}). If it were not the case, there would be a risk of having a lattice with a robust set that is too small for the desired payload. 

\subsection{Syndrome-Trellis Coding}\label{subsec:coding}

Lastly, we explain how to use the proposed methodology for practical embedding with Syndrome-Trellis Codes (STC).  Fig.~\ref{fig:trellis} shows Alice's action on a single lattice. Given $i$-th pseudo-stego, she will take the $i$-th lattice as a vector and inspect its recompressed values to asses the coefficients' robustness. She will then perform embedding on the recompressed lattice according to Section~\ref{subsec:latEmb}. However, she cannot simply send the output of the coding mechanism because the channel's recompression can potentially change the non-robust coefficients, which would prevent Bob from reading the secret message. Instead, Alice puts the original non-robust coefficients (before recompression) back into the lattice, which yields the $(i+1)$-th pseudo-stego. This way, it is ensured that after the recompression, Bob will be able to decode the secret message. 

For the actual STC implementation, we used a python wrapper of a C++ implementation, with the height constraint $h=10$.\footnote{\url{https://github.com/daniellerch/pySTC}} With this particular implementation, we observed an undesirable behavior when the message size is `too small' (\textit{e.g.} less than 0.5\% of the changeable elements), namely the codes would make an embedding change of magnitude 2. While this would not happen often, it would ultimately destroy the communicated message. Since this seems to be the STC implementation flaw, we discarded the experiments affected by this phenomenon. From a practical point of view, this is not an issue, as the steganographer can easily verify if the coding mechanism created such an embedding change and either embed the cover image again or discard this combination of image/message for transmission.

\begin{table*}[h]
    \begin{centering}
    \begin{tabular}{|c|c|c|c|c|c|c|c|c|c|}
    \hline 
    \textbf{Emb. rate } & \multirow{2}{*}{\textbf{Scan}} & \multirow{2}{*}{\textbf{Coder}} & \multirow{2}{*}{\textbf{94}} & \multirow{2}{*}{\textbf{95}} & \multirow{2}{*}{\textbf{96}} & \multirow{2}{*}{\textbf{97}} & \multirow{2}{*}{\textbf{98}} & \multirow{2}{*}{\textbf{99}} & \multirow{2}{*}{\textbf{100}}\tabularnewline
    \textbf{(bpnzAC)} &  &  &  &  &  &  &  & & \tabularnewline
    \hline 
    \hline 
    0.1 & all & all & 100 & 100 & 100 & 100 & 100 & 100 & 100\tabularnewline
    \hline 
    \hline 
    0.2 & Random and High-Low & all & 100 & 100 & 100 & 100 & 100 & 100 & 100\tabularnewline
    \hline 
    \hline 
    0.2 & Low-High & \texttt{convert}  & 100 & 100 & 99.98 & 99.72 & 94.53 & 89.95 & 99.97\tabularnewline
    \hline 
    0.2 & Low-High & \texttt{mozjpeg}, no optim  & 100 & 100 & 100 & 99.98 & 98.22 & 88.25 & 99.89\tabularnewline
    \hline 
    0.2 & Low-High & \texttt{mozjpeg}, optim  & 100 & 100 & 99.98 & 99.98 & 99.44 & 92.91 & 99.92\tabularnewline
    \hline 
    \hline 
    0.3 & Low-High & \texttt{convert}  & 100 & 99.13 & 99.05 & 71.18 & 44.03 & 25.09 & 30.07\tabularnewline
    \hline 
    0.3 & High-Low & \texttt{convert}  & 100 & 100 & 100 & 100 & 100 & 100 & 20.21\tabularnewline
    \hline 
    0.3 & Random & \texttt{convert}  & 100 & 100 & 100 & 100 & 100 & 87.09 & 27.12\tabularnewline
    \hline 
    \hline 
    0.3 & Low-High & \texttt{mozjpeg}, no optim  & 100 & 100 & 99.08 & 99.05 & 58.13 & 28.99 & 30.24\tabularnewline
    \hline 
    0.3 & High-Low & \texttt{mozjpeg}, no optim  & 100 & 100 & 100 & 100 & 100 & 100 & 19.93\tabularnewline
    \hline 
    0.3 & Random & \texttt{mozjpeg}, no optim  & 100 & 100 & 100 & 100 & 99.99 & 98.31 & 27.55\tabularnewline
    \hline 
    \hline 
    0.3 & Low-High & \texttt{mozjpeg}, optim  & 100 & 100 & 99.96 & 99.89 & 72.74 & 45.64 & 49.37\tabularnewline
    \hline 
    0.3 & High-Low & \texttt{mozjpeg}, optim  & 100 & 100 & 100 & 100 & 100 & 100 & 63.5\tabularnewline
    \hline 
    0.3 & Random & \texttt{mozjpeg}, optim  & 100 & 100 & 100 & 100 & 99.98 & 99.39 & 57.82\tabularnewline
    \hline 
    \hline 
    0.4 & Low-High & \texttt{convert}  & 100 & 99.95 & 84.53 & 34.9 & 18.8 & 10.18 & 7.96\tabularnewline
    \hline 
    0.4 & High-Low & \texttt{convert}  & 100 & 100 & 100 & 100 & 100 & 100 & 4.82\tabularnewline
    \hline 
    0.4 & Random & \texttt{convert}  & 100 & 100 & 100 & 100 & 99.98 & 33.12 & 6.65\tabularnewline
    \hline 
    \hline 
    0.4 & Low-High & \texttt{mozjpeg}, no optim  & 100 & 100 & 99.94 & 96.71 & 23.03 & 11.85 & 7.89\tabularnewline
    \hline 
    0.4 & High-Low & \texttt{mozjpeg}, no optim  & 100 & 100 & 100 & 100 & 100 & 100 & 4.83\tabularnewline
    \hline 
    0.4 & Random & \texttt{mozjpeg}, no optim  & 100 & 100 & 100 & 100 & 99.97 & 59.59 & 6.63\tabularnewline
    \hline 
    \hline 
    0.4 & Low-High & \texttt{mozjpeg}, optim  & 100 & 100 & 99.92 & 98.07 & 39.73 & 22.34 & 19.97\tabularnewline
    \hline 
    0.4 & High-Low & \texttt{mozjpeg}, optim  & 100 & 100 & 100 & 100 & 100 & 100 & 34.83\tabularnewline
    \hline 
    0.4 & Random & \texttt{mozjpeg}, optim  & 100 & 100 & 100 & 99.98 & 99.97 & 76.0 & 26.61\tabularnewline
    \hline 
    \end{tabular}
    \par\end{centering}
    \vspace{2mm}
    \caption{Embedding success rates for different JPEG quality factors (in \%) with \texttt{convert} and \texttt{mozjpeg} with and without rate-distortion optimization.}
    \label{tab:embfailure}
\end{table*}

\section{Results}\label{sec:results}

The goal of this section is to benchmark the different characteristics of the proposed scheme within the framework of robust steganography. The considered setup will be the same as the one depicted in Fig.~\ref{fig:setting} and recalled in Section~\ref{subsec:security}, \textit{i.e.}~the steganographer will embed its payload on a single-compressed image and submit it on the platform that will compress the stego image. On the other side, the steganalyst will observe contents that are published on the platform, \textit{i.e.}~contents which are double-compressed, either in their cover or stego version. The lossy coding algorithm and used parameters by the steganographer and the platform will be identical. The considered figures of merit are the following: 
\begin{itemize}
    \item The {\it practical security} of the robust scheme for different scanning strategies and JPEG compressors. It is also important to compare it w.r.t. the naive embedding which is not robust but maximizes the security and can be considered as a baseline. 
    \item The {\it embedding success rate}, \textit{i.e.} the probability to be able to embed the prescribed payload. The size of the robust set being limited, for high-quality factors, the robust set may be too small on several images.
    \item The {\it impact of the compressor}, such as \texttt{mozjpeg} or \texttt{convert} (used by the {\it Slack} application, see Section~\ref{subsec:slack}), which can use specific quantization matrices or can optimize the rate-distortion trade-off and which can change quantization values before applying Huffman coding.\\   
\end{itemize} 

These different features are evaluated using a classical steganography/steganalysis setup:\\
\begin{itemize}
    \item Images from BOSSBase~\cite{bossbase} in greyscale format are used as a source of covers. Images in the pixel/PGM format are used as pre-covers and then compressed with the appropriate compressor.
    \item Regarding steganography, UERD~\cite{guo2015using} is chosen as the embedding scheme because it offers a good tradeoff between complexity and practical security. Different payload sizes ranging from 0.1 bpnzAC to 0.4 bpnzAC are adopted.
    \item Regarding steganalysis, DCTR features~\cite{holub2015low} are combined with the regularized linear classifier~\cite{cogranne2015ensemble}. 5,000~pairs of images are used for training, and 5,000~pairs for testing.
    \item The classical probability of error $P_e$ minimizing the sum of false positive and false negative rates during training is reported as a measure of practical security.
\end{itemize} 
Different quality factors, ranging from 50 to 100, are reported to analyze the evolution of the mentioned features w.r.t.~JPEG quantization. Note, however, that there is a substantial discrepancy between the quantization matrices used by \texttt{libjpeg/convert} and \texttt{mozjpeg}, the quantization steps used by \texttt{mozjpeg} being always greater or equal to the quantization steps used by \texttt{convert}. 

Note also that, except when explicitly mentioned in the case of rate-distortion optimization, the proposed scheme is errorless. Consequently, no error-correcting codes need to be used, and no error transmission probability needs to be reported.

\subsection{Practical security}
Fig.~\ref{fig:pract_sec_convert}, Fig.~\ref{fig:pract_sec_mozjpegnotreillis} and Fig.\ref{fig:pract_sec_mozjpegtreillis} present respectively the detection error $P_e$ for respectively \texttt{convert}, \texttt{mozjpeg} without (option \texttt{-notrellis} added) and \texttt{mozjpeg} with the rate-distortion trade-off.

Several remarks can be drawn from this extensive set of experiments.

\paragraph{On the impact of the scan strategy}

The strategy of randomly picking the DCT modes for each lattice offers a gain of practical security w.r.t. to scans starting with low frequencies or high frequencies except for very high-quality factors (\textit{i.e.} $\geq95$). For quality factors below 85, the gain associated with the random scan is between 2 and 5\% w.r.t.~the other scans. However, one can choose the scan starting with low-frequency modes for high-quality factors. 
It is also interesting to notice that starting with low frequencies is, on average, a better strategy than scanning the high frequencies first, even if the size of the robust set is far larger with the second option (see Fig.~\ref{fig:rob_vs_lattice}). We can see that there is a tradeoff between the size of the robust set and the modes it considers as robust. 

On one side, the high-frequency modes are more robust since they are associated with bigger, hence more conservative, quantization steps; on the other side, they are more detectable. 

Note also that, on average, the random-scan strategy has to be preferred because of its higher security and its possibility to spread the payload without first computing the robust set (see Section~\ref{sec:embAlgo}). 

\paragraph{On the gap between robust and non-robust embedding}
When comparing with the most favorable embedding strategy, we can observe that, at 0.1 bpnzAC, the gap is very small (\textit{i.e.}~below 3\% in terms of detection error) but becomes substantial (\textit{i.e.} reaching more than 10\% for few quality factors) for larger embedding rates and high-quality factors. This is not surprising since we have seen in Section~\ref{sec:embAlgo} that the size of the robust set tends to decrease w.r.t. the quality factor, which means that the embedding algorithm has to perform more embedding changes. Indeed, for the same payload size, the number of embedding changes decreases w.r.t. the number of changeable coefficients. The larger detectability is also due to the fact that the non-robust coefficients initially associated with small embedding costs in a non-robust setting cannot be modified anymore with the proposed scheme. They are lost for embedding.

\paragraph{On the impact of the quality factor}

For the \texttt{convert} coder, we can notice a ``bump" for quality factors between 90 and 100, which is associated with an increase of detectability w.r.t. the non-robust scheme. For the \texttt{mozjpeg} coder, we can also observe oscillations of the overall detectability. We hypothesize that these two phenomena are due to the interplay between DCT modes quantized with specific steps. From Fig.~\ref{fig:robust_size}, we can observe that the oscillations in terms of detectability are on par with the ones coming from the size of the robust sets. Note also that part of the non-monotony of the detectability is also due to the quantization tables only and was already observed for plain JPEG steganography~\cite{butora2019effect}. 

\paragraph{On the impact of the coder}

If, as reported above, the different quantization tables used by the two coders are associated with different detectabilities, we can also notice that whenever the \texttt{mozjpeg} coder uses rate-distortion optimization, the practical detectability is smaller. If we noticed that the payload size is smaller (the number of 0s increases by about 10\% after the optimization), it is probably not the only reason since the same decrease of 0s is observed between \texttt{convert} and \texttt{mozjpeg} without optimization. 

We hypothesize that the produced cover images with optimization are also more ``secure" sources since they have less isolated non-zero modes due to the optimization process.    

\paragraph{On the impact of filtering}

\begin{figure}[ht]
	\begin{centering}
		\includegraphics[width=0.9\columnwidth]{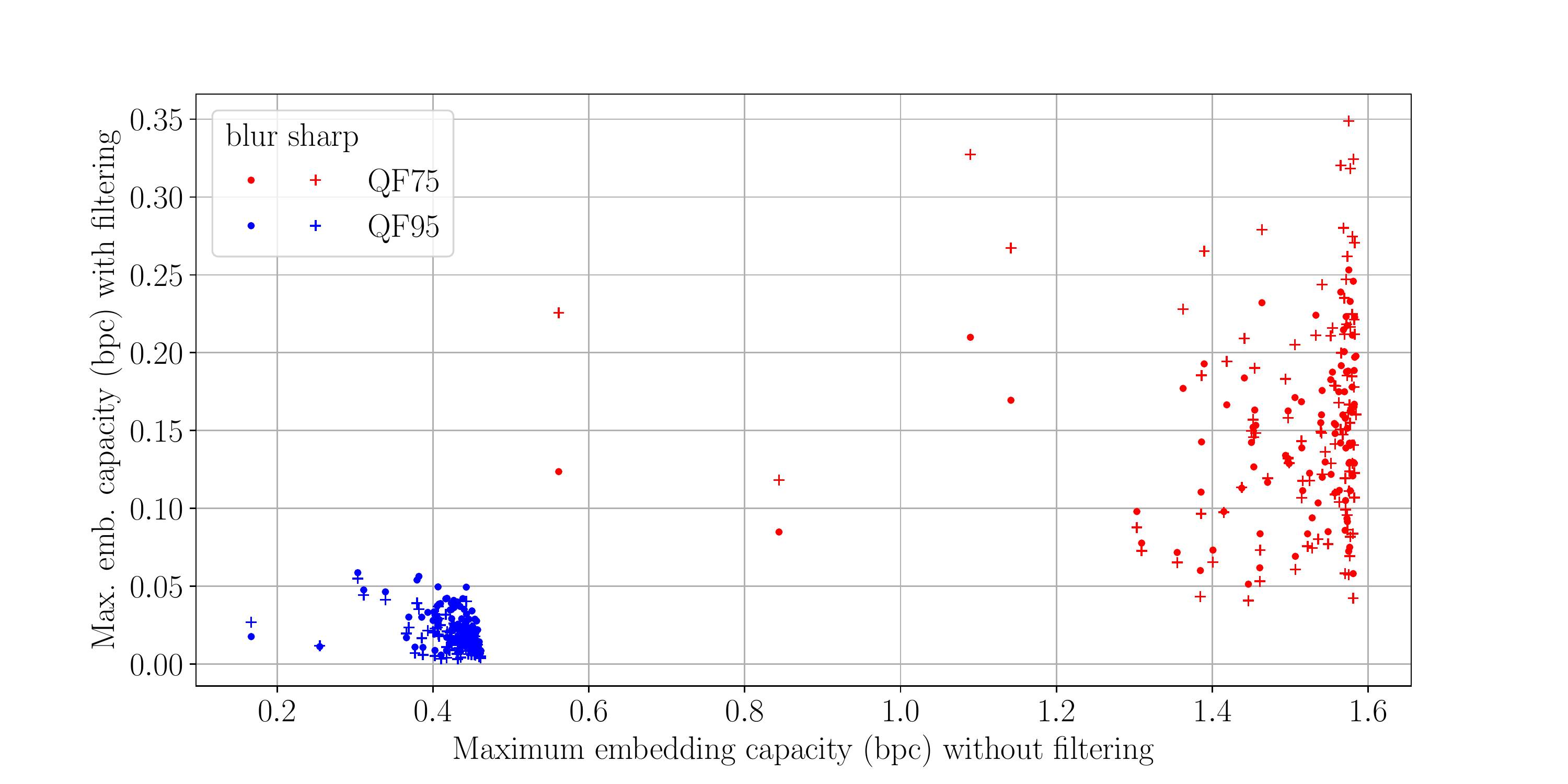}
		\caption{Maximum embedding capacity evolution with extra spatial filtering over 100 randomly selected images with \texttt{convert} and Low-High scanning strategy.\label{fig:rob_size_filter}}
		\par\end{centering}
\end{figure}

In Fig.~\ref{fig:rob_size_filter} we show how the maximum embedding capacity changes if we additionally process the decompressed image before recompression. We used \texttt{convert} at two different quality factors and a Low-High scanning strategy. Two operations are considered: $3\times3$~blurring with Gaussian kernel and $3\times3$ sharpening (both available in \texttt{convert}). The capacity was computed assuming optimal coder allowing to communicate $1$ bit per coefficient robust towards one embedding change and $\log_2 3$ bits per coefficient robust towards both embedding changes. We can see that both processing operations decrease the capacity from up to $\log_2 3$ bits per coefficient (bpc) to less than $0.35$ bpc at QF 75, and at QF 95, the maximum capacity decreases from $0.3-0.45$ bpc to less than $0.06$ bpc, preventing the sender from using bigger messages. Moreover, since the attainable payloads are so small, the security suffers as well, because the steganographer simply cannot commit to embedding changes associated with small embedding costs. At 0.1 bpnzAC, the probability of error is 3\% for QF 75 and 1\% for QF 95 respectively. However, despite increased detectability and limited embedding capacity, the robustness of the method is still guaranteed.

{\tiny }

\paragraph{Detectability with SOTA detectors}

Because of the increased complexity of Deep Learning steganalyzers, we provide only limited evaluation with JIN-SRNet~\cite{boroumand2018deep},\cite{butora2021JIN}. In Table~\ref{tab:srnet}, we show the detection error rates for the proposed scheme with a Low-High scanning strategy and original (non-robust, single-compressed) UERD. Interestingly enough, the proposed scheme is in most cases more secure than the original UERD in single-compressed images. We believe this is due to preventing some `easy-to-catch' embedding changes during the recompression process. Most importantly, we see that our robust methodology does not increase the detectability even with a state-of-the-art detector. 

\paragraph{Detectability before or after recompression}

As a side-experiment, we also performed steganalysis on Cover or Stego images before recompression (this would be equivalent to a practical scenario where Eve has access to inputs $(\mathbf{C_1},\mathbf{S_1})$ of the platform in Fig.~\ref{fig:setting}), and we did not notice any gap between the practical security of the scheme before or after recompression.

\begin{table}[h]
    \begin{centering}
    \begin{tabular}{|c|c|c|c|c|c|}
    \hline 
    \multirow{2}{*}{\textbf{Method}} & \multirow{2}{*}{\textbf{QF}} & \multicolumn{4}{c|}{\textbf{Emb. rate (bpnzAC)}} \tabularnewline
    \cline{3-6}
     &  & \textbf{0.1} &  \textbf{0.2} &  \textbf{0.3} & \textbf{0.4} \tabularnewline
    \hline 
    \hline 
     \multirow{2}{*}{\textbf{Proposed}}& 75 & 0.1392 & 0.0525 & 0.0195 & 0.0088\tabularnewline
    \cline{2-6} 
     & 95 & 0.2742 & 0.1420 & 0.0403 & 0.0085\tabularnewline
    \hline 
    \hline 
   \textbf{UERD}& 75 & 0.1155 & 0.0428 & 0.0197 & 0.0075\tabularnewline
    \cline{2-6} 
     (non-robust) & 95 & 0.2423 & 0.1195 & 0.0735 & 0.0330\tabularnewline     
    \hline 
    \end{tabular}
    \par\end{centering}
    \vspace{2mm}
    \caption{Detection error of SRNet with \texttt{convert} and Low-High scanning strategy.}
    \label{tab:srnet}
    \end{table}
    
    \begin{table}[h]
    \begin{centering}
    \begin{tabular}{|c|c|c|c|c|c|c|c|}
    \hline 
    \multirow{2}{*}{\textbf{Method}} & \multirow{2}{*}{\textbf{QF}} & \multicolumn{6}{c|}{\textbf{Emb. rate (bpnzAC)}} \tabularnewline
    \cline{3-8}
     &  & \textbf{0.1} &  \textbf{0.2} &  \textbf{0.3} & \textbf{0.4} & \textbf{0.5}  & \textbf{0.6}\tabularnewline
    \hline 
    \hline 
     \multirow{4}{*}{\textbf{Proposed}}& 75 & 100 & 100 & 100 & 100 & 100 & 100\tabularnewline
    \cline{2-8} 
      & 92 & 100 & 100 & 100 & 100 & 100 & 99\tabularnewline     
    \cline{2-8} 
     & 95 & 96 & 95 & 92 & 92 & 80 & 73\tabularnewline
    \cline{2-8} 
     & 100 & 0 & 0 & 0 & 0 & 0 & 0\tabularnewline
    \hline 
    \hline 
    \multirow{4}{*}{\textbf{\cite{ZENG22}}}& 75 & 0 & 0 & 0 & 0 & 0 & 0\tabularnewline
    \cline{2-8} 
     & 92 & 0 & 0 & 0 & 0 & 0 & 0\tabularnewline
    \cline{2-8} 
      & 95 & 0 & 0 & 0 & 0 & 0 & 0\tabularnewline     
    \cline{2-8} 
     & 100 & 0 & 0 & 0 & 0 & 0 & 0\tabularnewline
    \hline 
    \hline 
     \multirow{4}{*}{\textbf{\cite{WU22SSR}}} & 75 & 0 & 0 & 0 & 0 & 0 & 0\tabularnewline
    \cline{2-8} 
      & 92 & 0 & 0 & 0 & 0 & 0 & 0\tabularnewline
    \cline{2-8} 
      & 95 & 0 & 0 & 0 & 0 & 0 & 0\tabularnewline      
    \cline{2-8} 
     & 100 & 0 & 0 & 0 & 0 & 0 & 0\tabularnewline
    \hline 
    \end{tabular}
    \par\end{centering}
    \vspace{2mm}
    \caption{STC embedding success rates (in \%) for different embedding schemes with \texttt{convert} at Quality Factors 75, 92, 95, and 100. The proposed method was combined with the random scanning strategy. Height constraint of the STC was set to $h=10$.}
    \label{tab:STCfailure}
    \end{table}
    
\subsection{Embedding success rate}\label{subsec:success}
We analyze here when the embedding fails, \textit{i.e.} when the payload size is too small or too large to not modify only robust coefficients.
\subsubsection{Impact of a small robust set}
Because the size of the robust set can be limited, the maximum embedding capacity for ternary embedding can be smaller than $\log_2(3)\simeq 1.58$ bit per coefficient. If the prescribed embedding rate (chosen to be in bpnzAC) is larger than the maximum achievable rate, we consequently report an embedding failure. 
Table~\ref{tab:embfailure} reports the embedding success rate computed on the BOSSBase database for the different coders, different scanning strategies, embedding rates ranging from 0.1 bpnzAC to 0.4 bpnzAC, and different quantization factors. We can draw several conclusions:
\begin{itemize}
    \item For QF $\leq94$, the embedding success rate reaches 100\%, but the higher the quality factor, the less robust the embedding is. This is due to the fact that, in this range of quality factors, the size of the robust set decreases.
    \item The Low-High scan is less robust than the Random scan, which is less robust than the High-Low scan. This is coherent with the size of the robust sets, which follow the same trend (the robust set associated with the Low-High is smaller than the robust set associated with the Random scan, which is smaller than the one associated with the High-Low) plotted in Fig.~\ref{fig:robust_size}
    \item The \texttt{mozjeg} coder with rate-distortion optimization is more robust than the same coder without optimization, which is, in turn, more robust than the \texttt{convert} coder. Again, this is coherent with the hierarchy on the robust set size, plotted in Fig.~\ref{fig:robust_size}.
\end{itemize}

\subsubsection{Impact of STCs}
Since the above analysis is assuming optimal coding mechanism, we now investigate the effect of a practical coding scheme, the STCs. Table~\ref{tab:STCfailure} shows embedding success rate across different methods while using \texttt{convert} coder. We consider the embedding to be successful, only if the whole secret message can be retrieved from the recompressed image. We can see that our method starts failing only for high qualities and with increasing payload. This is due to the small robust set size, which in turn forces the (sub-optimal) STC to embed into a non-robust coefficient. This is in line with the sudden drop in robust set size for qualities above 92 depicted in Fig.~\ref{fig:robust_size}. Note that this could be already verified at the sender's side and the steganographer can thus avoid sending a non-robust stego image. This problem is inherently related to the use of STCs and the fact that for very low or very high payload sizes, wet costs can be used.




\subsection{Impact of rate-distortion strategies}

\begin{figure*}[h]
    \begin{centering}
    \includegraphics[width=1.0\textwidth]{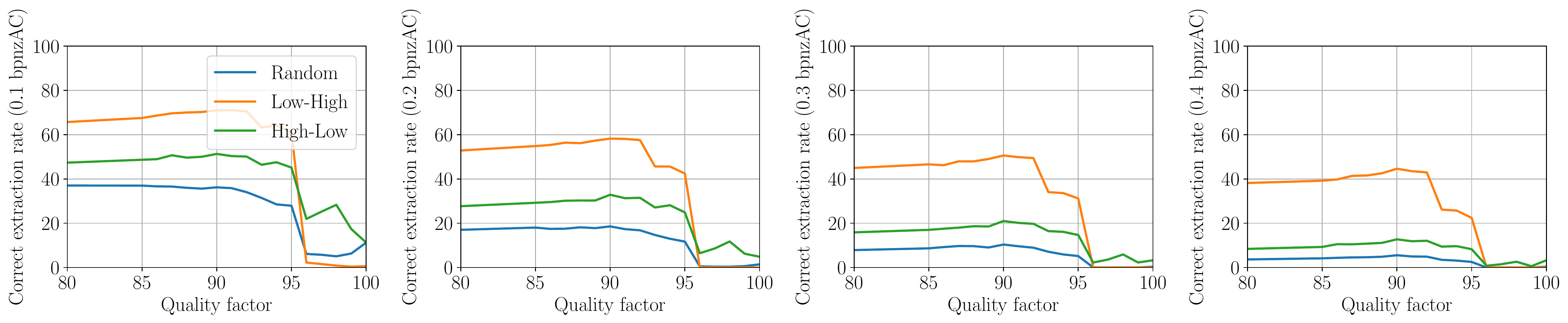}
    \caption{Correct extraction rate (in \%) for \texttt{mozjpeg} with rate-distortion optimization for different embedding rates.\label{fig:mozjpeg_treillis_fail2extract}}
    \par\end{centering}
\end{figure*}

This last experiment investigates to what extent the rate-distortion strategy, which is used by the \texttt{mozjpeg} coder to decrease the file size by changing DCT coefficients values and to decrease their associated Huffman code length, is detrimental to the robustness of the scheme. This is conducted out of curiosity since we know in advance that the proposed scheme is not robust to change of coefficient after quantization. Fig.~\ref{fig:mozjpeg_treillis_fail2extract} shows the ratio of images that can convey the payload when this option is activated.

Here we can see two behaviors:
\begin{enumerate}
    \item For quality factors $\leq 95$, starting with low-frequency coefficients increases the correct extraction rate significantly. For quality factors $>95$, starting with the High-frequency coefficient offers the best extraction rate. Moreover, there is an overall drop in extraction rate for quality factors $>95$, which is caused by reduced robust set size (see Fig.~\ref{fig:robust_size}).
    \item The larger the embedding rate, the smaller the number of images having a correct extraction. However, we can note that in a favorable setting (Low to High scan and QF below 95 at 0.1 bpnzAC), the ratio of the correctly extracted payload is larger than 60\%.  
\end{enumerate}

\subsection{Comparison with prior art}
In Table~\ref{tab:secComparison}, we compare the security of the proposed method at JPEG qualities 75 and 95 to those of MINICER~\cite{ZENG22} and SSR~\cite{WU22SSR}. To have a fair security comparison without any issues caused by STCs, we only used simulated embedding. Because the SSR method changes the sign of DCT coefficients during embedding, we see very small detection errors across all payloads for both quality factors. The MINICER, on the other hand, has a security performance comparable to the proposed method at QF 75 and is even more secure at the higher quality. We explain this by the fact that MINICER assigns wet costs to all the DC modes during its cost update. This is a detail worth commenting upon because we observed in our experiments that the DC mode is producing the most robust coefficients. We, therefore, believe that MINICER does this update simply for security reasons. We will not follow this logic because we are not designing a new steganographic scheme, but instead are giving a general methodology for robust steganography. To this end, our main objective is the robustness of a given embedding scheme.

Finally, we would like to point out that even though MINICER is less detectable than our method, Table~\ref{tab:STCfailure} shows that both \cite{ZENG22} and \cite{WU22SSR} are not robust in any of the studied cases, while our method is fully robust at the lowest quality and more than 90\% robust even at quality 95 for payloads below 0.5 bpnzac. This can be explained by studying the two methods in a little bit more detail. We learned that \cite{WU22SSR} relies on the fact that embedding changes do not change the sign of the DCT coefficients. It is practically not true because many embedding changes are performed on coefficients equal to $-1,0,1$. Consequently, after recompression, the sign of these coefficients can change with high probability. The method~\cite{ZENG22}, on the other hand, assumes that DCT coefficients are changed during recompression only in blocks containing saturation. This is however not the only case. We observed that steganographic embedding changes combined with recompression can also cause other coefficients from the same DCT block to change, which gravely affects the robustness of~\cite{ZENG22}.

Note that the scheme proposed in the paper does not have these drawbacks since, thanks to the lattice embedding, a coefficient can be modified if and only if all the previously modified coefficients do not change (see condition R1 in Section~\ref{subsec:embAlgo}). 

\begin{table}[h]
    \begin{centering}
    \begin{tabular}{|c|c|c|c|c|c|}
    \hline 
    \multirow{2}{*}{\textbf{Method}} & \multirow{2}{*}{\textbf{QF}} & \multicolumn{4}{c|}{\textbf{Emb. rate (bpnzAC)}} \tabularnewline
    \cline{3-6}
     &  & \textbf{0.1} &  \textbf{0.2} &  \textbf{0.3} & \textbf{0.4} \tabularnewline
    \hline 
    \hline 
     \multirow{2}{*}{\textbf{Proposed}}& 75 & 0.5000 & 0.3999 & 0.2365 & 0.1732\tabularnewline
    \cline{2-6} 
     & 95 & 0.4997 & 0.4583 & 0.2926 & 0.1386\tabularnewline
    \hline 
    \hline 
    \multirow{2}{*}{\textbf{\cite{ZENG22}}}& 75 & 0.4579 & 0.4050 & 0.2154 & 0.1340\tabularnewline
    \cline{2-6} 
     & 95 & 0.4999 & 0.4574 & 0.4030 & 0.3618\tabularnewline
    \hline 
    \hline 
     \multirow{2}{*}{\textbf{\cite{WU22SSR}}} & 75 & 0.0451 & 0.0047 & 0.0019 & 0.0025\tabularnewline
    \cline{2-6} 
      & 95 & 0.0140 & 0.0082 & 0.0076 & 0.0022\tabularnewline
    \hline 
    \end{tabular}
    \par\end{centering}
    \vspace{2mm}
    \caption{Detection error with \texttt{convert}. The proposed method was combined with the random scanning strategy.}
    \label{tab:secComparison}
\end{table}


\section{Conclusions and Perspectives}

In this work, we introduced a methodology for JPEG steganography robust against subsequent recompression. Although the JPEG compressor has to be assumed known, it does not present any obstacles because, in a typical channel, a social media, we can easily access the compressor. Moreover, we noticed that the recompression pipeline present in the professional social network {\it Slack} produces the very same results as the \texttt{convert} compressor, which validates this assumption. 

First, we introduced the notion of the robustness of a DCT coefficient. We showed that this could be done by dividing the image into 64 non-overlapping lattices and performing $64$ consecutive recompressions (associated with $\pm1$ modifications) of an image, one per lattice. We introduced three ordering of the lattices: Low to High, High to Low, and Random. We showed that these three strategies offer different robust sets. Then we combined the coefficients’ robustness with steganographic costs from a non-robust stego algorithm in a straightforward way to robustify the algorithm. Additionally, it was shown how this could be done in a practical setting with Syndrome-Trellis Codes.

In the last part of the paper, we evaluate the security of our method with machine-learning steganalysis. We observe that the security is affected by everything in the system: Quality Factor, compressor, and the scanning strategy of the lattices. We link the differences in security to different sizes of the robust sets. Moreover, we can observe security loss compared to the non-robust version of the stego algorithm, which is expected because many coefficients with small embedding costs will not be usable for robust embedding. Lastly, unlike any of the preceding works on robust steganography, our method is truly errorless, giving us guarantees on the readability of the embedded secret message. The only exception to this is \texttt{mozjpeg} which allows rate-distortion optimization. On the other hand, we have seen that, if successfully embedded, the rate-distortion optimization increases the security of the underlying scheme.

In the future, we plan to derive theoretical bounds on the embedding capacity in the noisy recompression channel. The source code of the proposed robust embedding is available from \url{https://janbutora.github.io/downloads/}.

\bibliographystyle{IEEEtran}
\bibliography{totaleBibDesk.bib}

\end{document}